\documentclass{aa}
\usepackage[dvips]{graphicx}
\newcommand{\gtrsim}{\mathrel{\hbox{\rlap{\hbox{\lower4pt\hbox{$\sim$}}}\hbox{$>$}}}}
\newcommand{\lesssim}{\mathrel{\hbox{\rlap{\hbox{\lower4pt\hbox{$\sim$}}}\hbox{$<$}}}}

\def\AMM{NH$_3$}

\def\HII{H{\sc ii} }

\def\UC{UC~H{\sc ii}}

\def\jy{~Jy~beam$^{-1}$}
\def\mjy{~mJy~beam$^{-1}$}

\begin{document}
\title{Search for massive protostar candidates in the southern hemisphere: II. Dust continuum
emission\thanks{Based on observations collected at the European Southern Observatory (ESO) using the 
Swedish-ESO Submillimetre Telescope (SEST), La Silla, Chile }} 
%\title{Search for massive protostar candidates in the southern hemisphere: II. Dust continuum %emission\thanks{Based on observations carried out at the European Southern Observatory (ESO), La Silla, %Chile.}} 
\author{M.\ T.\ Beltr\'an\inst{1, }\thanks{{\it Present address:} Departament d'Astronomia i Meteorologia, Universitat
de Barcelona, Av.\ Diagonal
647, E-08028, Barcelona, Catalunya, Spain} \and J.\ Brand\inst{2} \and R.\ Cesaroni \inst{1} \and F.\ Fontani\inst{1} \and S.\ Pezzuto\inst{3} \and L.\ Testi\inst{1} \and S. Molinari\inst{3}} 
\institute{
INAF-Osservatorio Astrofisico di Arcetri, Largo E. Fermi 5,
I-50125 Firenze, Italy
\and
INAF-Istituto di Radioastronomia, Via Gobetti 101, I-40129 Bologna, Italy
\and
INAF-Istituto di Fisica dello Spazio Interplanetario, Via Fosso del Cavaliere, I-00133 Roma, Italy}

\offprints{M. T. Beltr\'an, \email{mbeltran@am.ub.es}}
\date{Received date; accepted date}

%\markboth{Beltr\'an et al.: }{}
\titlerunning{Massive star formation}
\authorrunning{Beltr\'an et al.}

\abstract{In an ongoing effort to identify and study high-mass protostellar candidates we have observed in
various tracers a sample of 235 sources selected from the IRAS Point Source Catalog, mostly with $\delta <
-30 \degr$, with the SEST antenna at millimeter wavelengths. The sample contains  142 {\it Low} sources and 93
{\it High}, which are believed to be in different evolutionary stages. Both sub-samples have been studied in
detail by comparing their physical properties and morphologies. Massive dust clumps have been
detected in all but 8 regions, with usually more than one clump per region. The dust emission shows a variety
of complex morphologies, sometimes with multiple clumps forming filaments or clusters. The mean clump has a
linear size of $\sim 0.5$~pc, a mass of $\sim 320~M_\odot$ for a dust temperature $T_{\rm d} = 30$~K, an H$_2$ density of
$9.5\times10^5$~cm$^{-3}$, and a surface density of 0.4~g\,cm$^{-2}$. The median values are 0.4~pc, $102~M_\odot$, 
$4\times10^4$~cm$^{-3}$, and 0.14~g\,cm$^{-2}$, respectively.
The mean value of the luminosity-to-mass
ratio, $L/M \simeq 99~L_\odot/M_\odot$, suggests that the sources are in a young, pre-ultracompact \HII\ phase. We have
compared the millimeter continuum maps with images of the mid-IR MSX emission, and have discovered 95 massive
millimeter clumps non-MSX emitters, either diffuse or point-like, that are potential prestellar or precluster
cores. The physical properties of these clumps are similar to those of the others, apart from the mass that is
$\sim 3$ times lower than for clumps with MSX counterpart. Such a difference could be due to the potential
prestellar clumps having a lower dust temperature. The mass spectrum of the clumps with masses above $M \sim
100~M_\odot$ is best fitted with a power-law $dN/dM \propto M^{-\alpha}$ with  $\alpha = 2.1$, consistent with the Salpeter~(\cite{salpeter55})
 stellar IMF, with $\alpha=2.35$. On the other hand,
the mass function of clumps with masses $10~M_\odot\lesssim M  \lesssim 120~M_\odot$ is better fitted with a
power law of slope $\alpha = 1.5$, more consistent with the mass function of molecular clouds derived from gas observations. 
\keywords{Stars: circumstellar matter -- stars: formation -- ISM: clouds -- radio continuum: ISM -- infrared: ISM}
}

\maketitle

\section{Introduction}

Massive stars ($M \geq 8~M_\odot$) play a crucial role in the appearance and evolution of galaxies.
They are responsible for the production of heavy elements and influence the interstellar medium
through energetic winds and supernovae. Despite their importance, the understanding of massive
star formation has remained significantly behind that of their lower-mass counterparts, for which
much observational and theoretical work has already been done. This situation has changed in recent
years, when the formation of high-mass stars has been gaining increasing interest, with the attention
gradually shifting from the study of clouds associated with ultracompact \UC\ regions, to those where
only luminous IRAS sources without radio continuum emission were detected. This is equivalent to
approaching the earliest stages of high-mass star formation, when most of the luminosity of the newly formed
(proto)star is derived from the release of gravitational energy.

Palla et al.~(\cite{palla91}) have used the IRAS Point Source Catalog (IRAS-PSC) to select plausible
candidates of massive (proto)stars with $\delta \geq -30\degr$. The basic criteria used by these authors
to select the targets are based on the IRAS colours and the lack of association with \HII regions: the
former constraint is derived from the study of Richards et al.~(\cite{richards87}), who used IRAS
colours to identify compact molecular clouds; the latter is aimed at biasing the sample towards
high-mass Young Stellar Objects (YSOs) in a very early phase of their evolution, when an \HII region has
not yet developed. The resulting sample was split into two sub-samples, this time using the IRAS colour
selection criteria of Wood \& Churchwell~(\cite{wood89}) for identifying \UC\ regions: $[25-12]\geq0.57$
and $[60-12]\geq1.30$. The two sub-samples of compact molecular clouds satisfying and non-satisfying the
Wood \& Churchwell criteria have been called {\it High} and {\it Low}, respectively. The selected
sources also verify that there are no upper limits for their fluxes at 25, 60, and 100~$\mu$m, and that 
$F_{60\mu{\rm m}}\geq 100$~Jy. Palla et al.~(\cite{palla91}) have searched for H$_2$O maser emission
associated with the sources and found a lower association rate for the {\it Low} sources  that has been
interpreted as an indication that the {\it Low} sources are in an earlier evolutionary phase than the
{\it High} sources. Both sub-samples have been systematically observed in different continuum and
molecular line tracers, from centimeter to near-infrared wavelengths (Molinari et al.~\cite{moli96},
\cite{moli98a}, \cite{moli98b}, \cite{moli00}, \cite{moli02}; Brand et al.~\cite{brand01}; Zhang et
al.~\cite{zhang01}, \cite{zhang05}). The scope was to derive the physical properties of the two groups
and confirm  that they are in different evolutionary stages. In particular, the {\it Low} sub-sample
would contain a fraction of young sources that are not yet zero-age main sequence (ZAMS) massive stars.
The main findings of this study have been thoroughly discussed by  Brand et al.~(\cite{brand01}) and can
be summarized as follow:  \begin{itemize}

\item {\it High} and {\it Low}  sources have luminosities typical of high-mass YSOs ($L>10^3L_\odot$);

\item the percentage of {\it Low}  sources not associated with \UC\ regions is higher (76\%; see Molinari et
al.~\cite{moli98a}) than that of {\it High}  sources (57\%);

\item a large fraction of {\it Low} sources have dust temperatures of $\sim 30$~K (Molinari et al.~\cite{moli00}) much lower than those measured towards ``hot cores'' ($\gtrsim 100$~K), and are hence most likely high-mass objects too young to have yet developed an \UC\ region.

\end{itemize}

In order to extend our systematic search for massive protostellar candidates to the entire sky, we have selected new targets in
the southern hemisphere, namely {\it High} and {\it Low} sources with $\delta < -30\degr$. This is the second of a series of papers
aimed to conduct in the southern hemisphere the same kind of investigation carried out for sources with $\delta \geq -30\degr$. In the first
paper  (Fontani et al.~\cite{fontani05}) we have discussed the results of the molecular line  survey towards the {\it Low}
sources with  $\delta < -30\degr$. The main results of that study are that there is a tight association of the sources with
dense gas, and that the physical properties of the {\it Low} sources, such as linewidths, and distribution of the NRAO VLA Sky
Survey NVSS-to-IRAS flux ratio, are comparable to those found by Sridharan et al.~(\cite{srid02}) for a sample of {\it
High}-like sources when the luminosity of the sources is $L < 10^5 L_\odot$. The mass-luminosity ratios are also similar to
those found by Sridharan et al.~(\cite{srid02}) but lower than the ratio found for a sample of \UC\ regions, supporting the idea
that our {\it Low} sources, as well as those {\it High}-like of Sridharan et al.~(\cite{srid02}), are younger than \UC\ regions.
In the present paper we discuss the main findings of the millimeter continuum survey carried out towards a sample of {\it High}
and {\it Low} sources with $\delta < -30\degr$, plus a few additional sources with $-30\degr < \delta < 40\degr$. In particular
we study the morphology of the dust emission and derive the physical properties of the sources, and  compare them with those
reported by other surveys in the literature. In addition, we also compare the properties of both {\it High} and {\it Low}
sub-samples, as well as those of millimeter sources associated with mid-infrared sources from the Midcourse Space Experiment
(MSX\footnote{MSX images have been taken from the on-line MSX database {\tt http://www.ipac.caltech.edu/ipac/msx/msx.html}})
Point Source Catalog and those that are not. Finally, we present a study on the mass spectrum of the observed sources.

\section{Sample}

The first step in extending the search for massive YSOs towards the southern hemisphere  was to select a sample
of possible candidates from the IRAS-PSC following the Palla et al.~(\cite{palla91}) criteria. Taking into
account the interest of studying the earliest stages of high-mass star formation, we first selected a sample of
{\it Low} sources with $\delta < -30\degr$, which was observed in  C$^{17}$O and/or CS and the results are presented
in the first paper by Fontani et al.~(\cite{fontani05}). Out of the 131 sources of this sample, 125 were then observed
in the continuum at millimeter wavelengths (the results of that survey are presented in this paper). In order to
conduct a comparative study of their properties with those of possibly more evolved sources, we also observed in
the continuum a comparable sample of {\it High} sources with $\delta < -30\degr$, which had been previously
detected in CS by Bronfman et al.~(\cite{bronfman96}).  In addition we also observed a number of {\it Low} and {\it
High} sources with $\delta > -30\degr$. The resulting sample observed in the millimeter continuum contains a
total of 235 sources: 142 of them {\it Low} and 93 {\it High}. Table~\ref{table_lumi} shows the position,
distance, and luminosity of the sources in the sample. As already mentioned, almost all (89\%) the sources in our
sample have $\delta < -30\degr$, whereas the Palla et al.~(\cite{palla91}) sample contained only objects
with $\delta \geq -30\degr$. Therefore, one may reasonably expect that our sample contains a higher
contamination by \HII or \UC\ regions than that of Palla et al.~(\cite{palla91}), because radio continuum
surveys of \HII regions with $\delta < -30\degr$ are less numerous than and not as complete as those with $\delta \geq
-30\degr$. 

All the {\it High}  sources in the sample have been previously detected in CS by Bronfman et
al.~(\cite{bronfman96}), and  those with $\delta > -30\degr$, with the exception of IRAS~18198$-$1429, also in
\AMM\ by Molinari et al.~(\cite{moli96}). The {\it Low} sources with $\delta < -30\degr$, with the exception of
IRAS~15579$-$5347, have been observed in CS by Fontani et al.~(\cite{fontani05}), and some of them have also been
observed in C$^{17}$O, and those with $\delta > -30\degr$ in \AMM\ by Molinari et al.~(\cite{moli96}).  Fifteen of the
{\it Low} sources with $\delta < -30\degr$ were not detected either in C$^{17}$O or CS (Fontani et
al.~\cite{fontani05}).

\section{Observations}
\label{obs}

The 1.2-mm continuum observations were carried out with the 37-channel bolometer array SIMBA (SEST Imaging
Bolometer Array) at the SEST (Swedish-ESO Submillimetre Telescope), on July 16--20, 2002 and July 9--13, 2003.

Maps of $900\arcsec \times 400\arcsec$ (azimuth $\times$ elevation) around each IRAS source in Table~\ref{table_lumi} were obtained,
with a scan rate of $80\arcsec/{\rm s}$, and a separation of $8''$ in elevation between scans. Bigger maps were also obtained by
using a mosaicing technique towards sources 16153$-$5016, 17225$-$3426, and 18014$-$2428 (here and in the following we will refer to
the sources without mentioning ``IRAS'' in the name). For the region surrounding source 16428$-$4109, the center of the observations
was shifted $\sim 4\farcm7$ northeast from the nominal IRAS point source position: after taking a first map centered at the IRAS
position, no millimeter emission was detected at the source nominal position but there clearly was emission at the edge of the map,
so we decided to shift the center towards the northeast. We have checked the HIgh RESolution (HIRES) IRAS images and have found
that there is an infrared source at the  nominal IRAS point source position, which indicates that the position given in the IRAS-PSC
is correct. In our study we have considered as associated with the IRAS source all the millimeter clumps located at $< 90''$ from
the nominal IRAS point source position. Therefore, the clumps detected towards 16428$-$4109 are not associated with the IRAS source.
The total integration time per map was about 15~minutes, and the typical noise level in the maps is 25--40\mjy. Atmospheric opacity
was determined from skydips, which were taken every 2 hours; values at zenith ranged between 0.21 and 0.50 (during the 2002
observations) and 0.13 and 0.30 (in 2003). The data were calibrated using observations of Uranus, made once or twice per day; the
conversion factor ranged between 58 and 75~mJy/count in 2002, and between 50 and 69 in 2003. The calibration uncertainty is  about
15\%. The pointing of the SEST was determined to be accurate within a few arcsec by observing the strong continuum sources $\eta$
Carinae or Centaurus A every 2 hours. The HPBW is $\sim 24''$.

All data were reduced with the program MOPSI, written by R.\ Zylka (IRAM, Grenoble), according to the
instructions given in the SIMBA Observers Handbook (2003). See Chini et al.~(\cite{chini03}) for a detailed
description of the data reduction method.

\section{Results}

\subsection{Kinematic distances}
\label{kindis}

Table~\ref{table_lumi} lists the kinematic distances, $d$, to the IRAS sources. The distances have been estimated, using the
rotation curve of Brand \& Blitz~(\cite{brand93}), from the CS line velocity given by Bronfman et al.~(\cite{bronfman96}) for
all the {\it High} sources, from the CS line velocity given by Fontani et al.~(\cite{fontani05}) for those {\it Low} with
$\delta < -30\degr$, and from the \AMM\ line velocity given by Molinari et al.~(\cite{moli96}) for the {\it Low} sources with
$\delta > -30\degr$. This method is valid for galactocentric distances between 2 and 25 kpc. For sources inside the solar
circle, there are two solutions for the kinematic distance, near and far. In some cases this ambiguity can be solved, for
example, when the height of the source from the galactic plane is more than 200~pc (i.e.\ roughly twice the scale height
of the molecular disk) at the far distance, or when the  near distance is too small for a 
massive star forming region. This is the case
for sources 14394$-$6004 and 17040$-$3959, for which $d_{\rm near}$ is $<100$~pc, and therefore the far distance was
adopted. There are many sources for which it was not possible to solve the distance ambiguity: for these in
Table~\ref{table_lumi} we report both distance estimates. {\it Note that in the following to derive the physical parameters of
the sources and in case of unsolved distance ambiguity, the near distance is adopted.} The distances to the IRAS sources are
between 130~pc and 27.1~kpc, with an average value of $\sim 4.5$~kpc, and a median value of $\sim 3.8$~kpc. 

We have not been able to derive an estimate of the distance for sources 08488$-$4457, 10088$-$5730,
10156$-$5804, 10575$-$5844, 11431$-$6516, 11476$-$6435, 12434$-$6355, 13078$-$6247, 13558$-$6159,
14198$-$6115, 14412$-$5948, 15506$-$5325, 15579$-$5347, 16204$-$4943, 16581$-$4212, 17140$-$3747,
17230$-$3531, 17231$-$3520, 17242$-$3513, 17352$-$3153, 17410$-$3019, 17425$-$3017. There are three
possible reasons for that: no estimate of the systemic velocity of the region was available; the
corresponding distances estimates are outside the galactocentric interval 2--25~kpc; the height from the
galactic plane is too large ($>200$~pc) for near and far distances.

Most of the regions observed contain more than one millimeter clump (see Section~\ref{clumpfind}), and it is
likely that all of them belong to the same star forming region as the IRAS source. Therefore, when deriving their
physical parameters, we have made the assumption that all the clumps are located at the same kinematic distance as
the one estimated for the IRAS source.

\subsection{Luminosities}
\label{lumi}

\begin{figure}
\centerline{\includegraphics[angle=0,width=6.5cm]{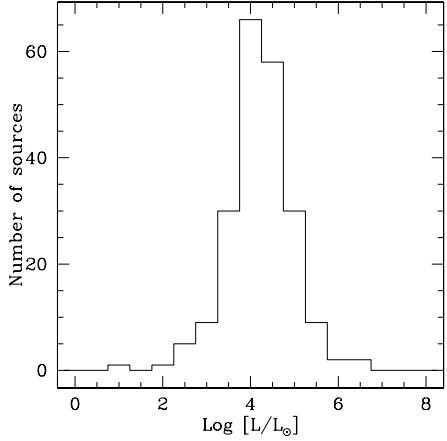}}
%\vspace*{-3cm}
\caption{Histogram of the luminosity distribution calculated from the IRAS flux densities of the sources.} 
\label{histo_1}
\end{figure}

The bolometric luminosities in  Table~\ref{table_lumi} were calculated by integrating the IRAS flux densities. The contribution from
longer wavelengths was taken into account by extrapolating according to a black-body function that peaks at 100~$\mu$m and has the
same flux density as the source at this wavelength.  The distribution of luminosities is shown in Fig.~\ref{histo_1}. The average
luminosity is $\sim 6.7\times10^4 L_\odot$ and the median is $\sim 1.6\times10^4 L_\odot$. These luminosities are $\sim 3.5$ times
lower than the average and median values derived by Fa\'undez et al.~(\cite{faundez04}) for their sample of southern sources. This
could be due to the fact that their sample has been selected from the survey of Bronfman et al.~(\cite{bronfman96}), which contains
high-mass very YSOs but it is also contaminated by more evolved sources such as \HII regions (or equivalently more massive stars),
which are expected to be brighter at far-infrared wavelengths. 

%If one takes into account the millimeter
%fluxes at 1.2~mm, then the derived luminosities could be $\sim$ 30\% lower in some cases.

%figure 1

%figure 3
\addtocounter{figure}{+1}
\begin{figure}
\centerline{\includegraphics[angle=0,width=8cm]{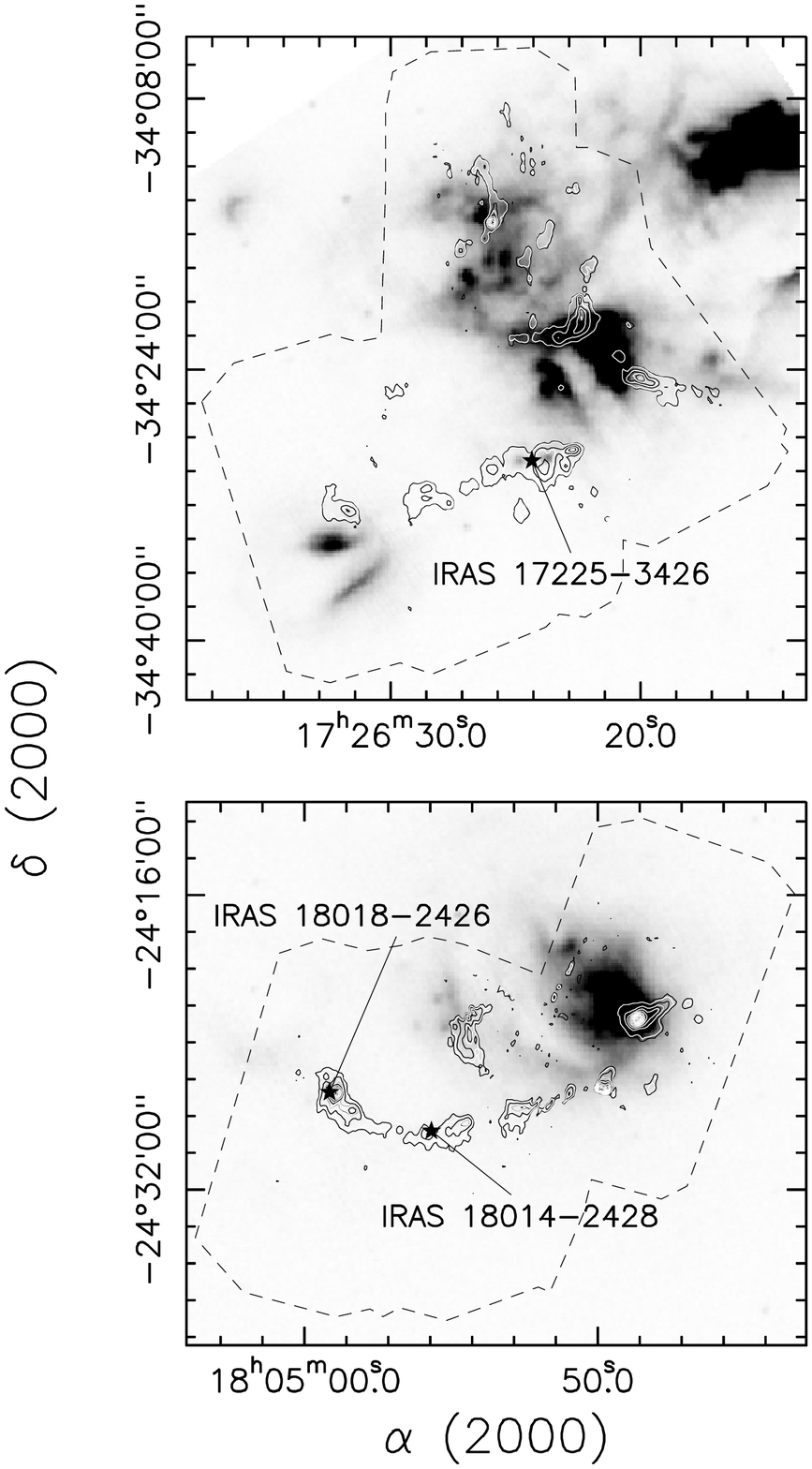}}
\caption{Overlay of the 1.2~mm continuum emission ({\it contours}) obtained with SIMBA at the SEST antenna, on the MSX emission at 21~$\mu$m
(Band E) ({\it image}) towards the sources IRAS~17225$-$3426 ({\it upper panel}) and IRAS~18014$-$2428 ({\it bottom panel}). The contour levels
range from 0.11 (3$\sigma$) to 1.67\jy\ in steps of 0.22\jy\ ({\it upper panel}) and from 0.09 (3$\sigma$) to 3.20\jy\ in steps of 0.17\jy\
({\it bottom panel}). The star marks the position of the IRAS sources.  The coordinates are in J2000 epoch. In the bottom panel, the two IRAS
sources correspond to Mol~36 (IRAS~18014$-$2428) and Mol~37 (IRAS~18018$-$2426) from Molinari et al.~(\cite{moli96}). The dashed polygon shows
the limits of the area mapped with SIMBA.}
\label{mosaic}
\end{figure}

\subsection{Identification of the clumps}
\label{clumpfind}

Massive dust clumps have been detected in all but 7 regions, 10102$-$5706, 11476$-$6435, 14198$-$6115,
15571$-$5218, 16403$-$4614, 16417$-$4445, and 18024$-$2231, of our sample. As already mentioned in
Section~\ref{obs}, for 16428$-$4109, the clumps are displaced by $\sim 4\farcm7$ from the position of the IRAS
source. The millimeter maps of our survey are shown in Fig.~\ref{maps_mm}. Figure~\ref{mosaic} shows the two
largest areas mapped in our survey, that is, the mosaiced maps towards sources 17225$-$3426 and 18014$-$2428. As
can be seen in the maps, there is usually  more than one clump  per region. In some cases the large number of
clumps and the extended emission detected in the region make it very difficult to separate them from each other
by eye. Therefore, in order to identify the millimeter clumps and their properties adopting a more objective
criterion, we have used a two-dimensional variation of the clump-finding algorithm {\it Clumpfind} developed by
Williams et al.~(\cite{williams94}). The three-dimensional version of the algorithm was originally designed to
be applied to spectral line datacubes, and the two-dimensional version is a simple modification of it. The
algorithm works by effectively contouring the data at a multiple of the rms noise of the map, then searching for
peaks of emission to locate the clumps, and finally following the clump profile down to lower intensities. The contouring
levels have to be chosen by hand, which means that the clump-finding procedure is not completely automated, and
therefore one can introduce biases into the results. {\it Clumpfind} does not {\it a priori} require any
particular shape of the clump profile, as some other clumps finding algorithms do, and one of its disadvantages
is that it misses low-mass clumps that lie below the lowest contour. This could flatten the low-mass end of the
mass spectrum of the regions. In our case, we set the lowest contour level to 3$\sigma$, and then increased the
contouring by steps of 3$\sigma$.

\begin{figure}
\centerline{\includegraphics[angle=0,width=11.5cm]{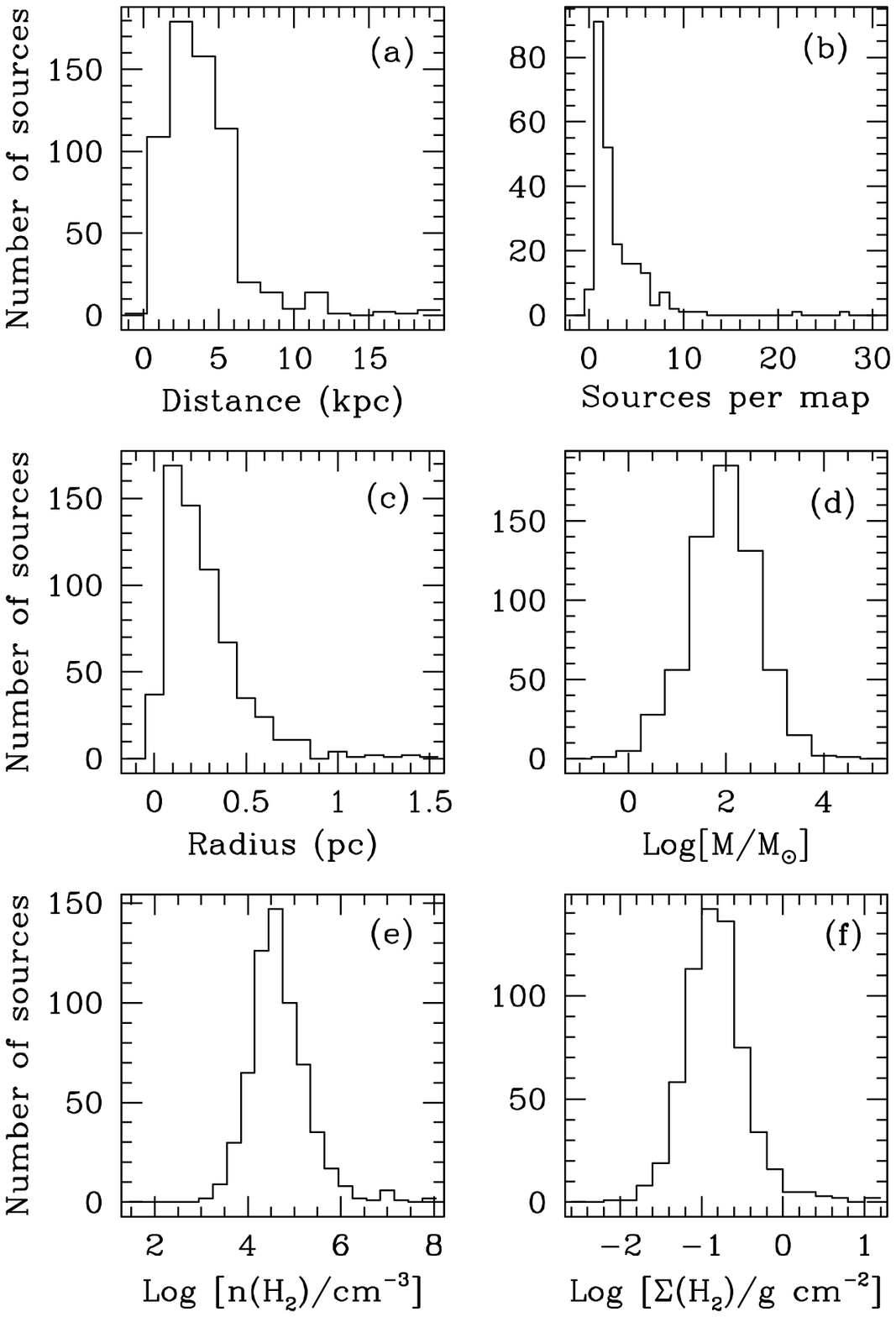}}
\caption{Histograms of some parameters of the clumps detected in the continuum emission at 1.2~mm:  {\it (a)} kinematic distance, 
{\it (b)} number of clumps per map, {\it (c)} radius of the clumps, {\it (d)} mass of the clumps,  {\it (e)} H$_2$ volume density,
and {\it (f)} H$_2$ surface density. Note that in the regions for which  mosaiced observations were carried out, we have included
all the clumps in the field when computing the histograms, and not only the sources in a field $900''\times 400''$, like for the
other regions.}
\label{histo_4}
\end{figure}

The clump-finding procedure calculates the peak position, the full width at half maximum (FWHM)  not corrected for beam size for the
x-axis, ${\rm FWHM_x}$, and for the y-axis, ${\rm FWHM_y}$, and the total flux density integrated within the clump boundary, that is
within the lowest contour level. Table~\ref{table_clumps} gives the offset positions in arcsec with respect to the nominal IRAS
point source position for each clump, the angular diameter, which has been calculated as the deconvolved geometric mean of ${\rm
FWHM_x}$ and ${\rm FWHM_y}$, the linear diameter, the total flux density, the mass, and the density. The last column indicates
whether the clump is associated with MSX emission, either point-like or diffuse, or not. Figures~\ref{histo_4}a and \ref{histo_4}b
show the histograms with the kinematic distance of the clumps  and  the number of clumps per region. The mean and median values of
the distance of the clumps are 3.9~kpc and 3.4~kpc, respectively, while those of the number of sources per region are 2.8 and 2.0,
respectively. Note that the values of the mean and median distance are slightly lower when taking into account all the clumps in the
regions instead of taking only into account the central IRAS source (see Section~\ref{kindis}). The maximum number of clumps have
been detected in the large mosaiced maps around sources 17225$-$3426 and 18014$-$2428, for which we have found 27 and 22 clumps,
respectively (see Fig.~\ref{mosaic}).

%figure 4

\subsection{Morphologies of the clumps}
\label{morpho}

The 1.2~mm maps in Fig.~\ref{maps_mm} show that the dust emission towards massive star forming regions presents a  variety of
complex morphologies. The emission, which usually peaks at or near the IRAS position, is very often associated with multiple
compact embedded objects and extended emission. Sometimes the emission is very centrally peaked, with a single massive clump
associated with the IRAS source, e.g.\ 12063$-$6259, 15454$-$5335, and 18144$-$1723, sometimes with more than one massive clump
clustered towards the central position, e.g.\ 12127$-$6244, 13333$-$6234, and 17149$-$3916, or with smaller and fainter clumps
clustered or located nearby, e.g.\ 15015$-$5720, 15519$-$5430, 17082$-$4114, and 17118$-$3090. Sometimes the emission shows
multiple clumps located throughout the field, which can be separated from each other, e.g.\ 16085$-$5138, 16093$-$5128,
16153$-$5016, and 17242$-$3513, or linked in a chain of clumps and elongated structures, e.g.\ 10184$-$5748, 13039$-$6108,
14000$-$6104, and 16164$-$4929, suggesting star formation in a sheet or a filament (e.g.\ Larson~\cite{larson85}). Particularly
nice examples of this latter phenomenon are seen in the mosaiced maps of 17225$-$3426 and 18014$-$2428 (Fig.~\ref{mosaic}) that show strings of
emission clumps and elongated and clustered structures. Sometimes the emission towards the IRAS position is very faint (e.g.\
08488$-$4457, 10277$-$5730, and 15464$-$5445), or no millimeter emission is  detected at all. In particular, the number of IRAS
sources not detected at 1.2~mm is 12, taking into account 10545$-$6244, 16153$-$5016, 16428$-$4109, and 17156$-$3607 as well,
for which although there are millimeter clumps detected in the region, none of them is associated (i.e.\ within $90''$)  with the IRAS source. All of them
are ${\it Low}$ sources. As can be seen in Table~3 of Fontani et al.~(\cite{fontani05}), 5 of them, 16403$-$4614, 16417$-$4445,
16153$-$5016, 16428$-$4109, and 17156$-$3607 have been detected in CS and C$^{17}$O, 3 of them, 10102$-$5706, 15571$-$5218, and
10545$-$6244 have been detected in CS but not in C$^{17}$O, and 3 of them, 11476$-$6435, 14198$-$6115, 14412$-$5948, have not
been detected in CS but not been observed in C$^{17}$O. Finally, one of the sources, 18024$-$2231, has been detected in NH$_3$
by Molinari et al.~(\cite{moli96}). On the other hand, there are 6 {\it Low} sources, 08488$-$4457, 10088$-$5730, 12434$-$6355,
15506$-$5325, 16204$-$4943, and 16581$-$4212, that have been detected in the millimeter continuum, although the emission is
faint, but not in CS (Fontani et al.~\cite{fontani05}), and one source, 15579$-$5347, that has not been observed in CS and has
been clearly detected in the continuum.

\subsection{Linear Diameters}

The deconvolved linear diameters of the clumps have been computed from their angular diameters. As mentioned in
Sect.~\ref{clumpfind}, both linear and angular diameters are listed in Table~\ref{table_clumps}. The mean value is $0.5$~pc for
those sources that have been resolved, which is in agreement with the average value of 0.6~pc found by Williams et
al.~(\cite{williams04}) for the sources of the Sridharan/Beuther sample (Sridharan et al.~\cite{srid02}; Beuther et
al.~\cite{beuther02}), and quite smaller than the value of 0.8~pc found by Fa\'undez et al.~(\cite{faundez04}), or 1~pc found by
Hill et al.~(\cite{hill05})  for their samples of southern sources, although it is not clear whether the sizes given by these
authors have been deconvolved or not. The median value for the clumps in our sample is 0.4~pc. Figure~\ref{histo_4}c shows the
histogram of the radius of the clumps.

We have searched for possible asymmetries in the clumps by calculating ${\rm FWHM_x}/{\rm FWHM_y}$. The mean and median values obtained, 1.04
and 0.96, respectively, indicate that the clumps are quite symmetric.

\subsection{Masses and densities}
\label{mass}

The masses of the clumps given in Table~\ref{table_clumps} have been estimated assuming that the dust emission is
optically thin, by using 
\begin{equation}
M_{\rm clump}=\frac{g\,S_\nu\,d^2}{\kappa_\nu\,B_\nu(T_{\rm d})},
\end{equation}
where ${S_\nu}$ is the flux density, $d$ is the distance to the source, ${\kappa_\nu}$ is the dust mass opacity
coefficient, $g$ is the gas-to-dust ratio, and $B_\nu(T_{\rm d})$ is the Planck function for a blackbody of dust
temperature $T_{\rm d}$, all measured at $\nu=250$~GHz. We adopted $\kappa_{250}$ = 1~cm$^2$\,g$^{-1}$ (Ossenkopf
\& Henning~\cite{ossenkopf94}),  $g=100$, and $T_{\rm d}= 30$~K for all the sources. Estimates of  $T_{\rm
d}$ have been obtained by fitting grey-bodies to the spectral energy distribution (SED) of  those IRAS sources
that have only one millimeter clump associated with them. As already done by Fontani et al.~(\cite{fontani05})
for the {\it Low}  sources in our sample, we have neglected the IRAS fluxes at 12 and 25~$\mu$m in the fit. The
reason for this choice is that two components are known to be present in the SEDs of luminous YSOs (see Sridharan
et al.~\cite{srid02} and Beuther et al.~\cite{beuther02}): one associated with compact, hot gas and dominating
the 12 and 25~$\mu$m fluxes; the other due to parsec-scale, colder material, contributing to the 60 and 100~$\mu$m
emission. The latter component is the one of interest to us, because we want to estimate the temperature of the
parcsec-scale clumps mapped at 1.2~mm. Adopting a dust absorption coefficient proportional to $\nu^2$, we
have found best fits with mean temperatures of 28 K for both {\it Low} and {\it High} sources. Therefore, taking into account
that for most of our clumps there are not enough measurements at different wavelengths to properly fit their SEDs
(for some of them the 1.2~mm is the only one available), we have decided to adopt $T_{\rm d}=30$~K for all the 
clumps. A similar value, $T_{\rm d}=32\pm5$~K, has been found by Molinari et al.~(\cite{moli00}) for a sample of
30 luminous {\it Low} sources in the northern hemisphere. Furthermore, Fa\'undez et al.~(\cite{faundez04}), who
have fitted the SED of a sample of southern hemisphere sources similar to the {\it High} sources  in our sample
with grey-bodies with two components, have found an average value $T_{\rm d}\simeq 32$~K for the colder
component.

%figure 5
\begin{figure}
\centerline{\includegraphics[angle=0,width=11cm]{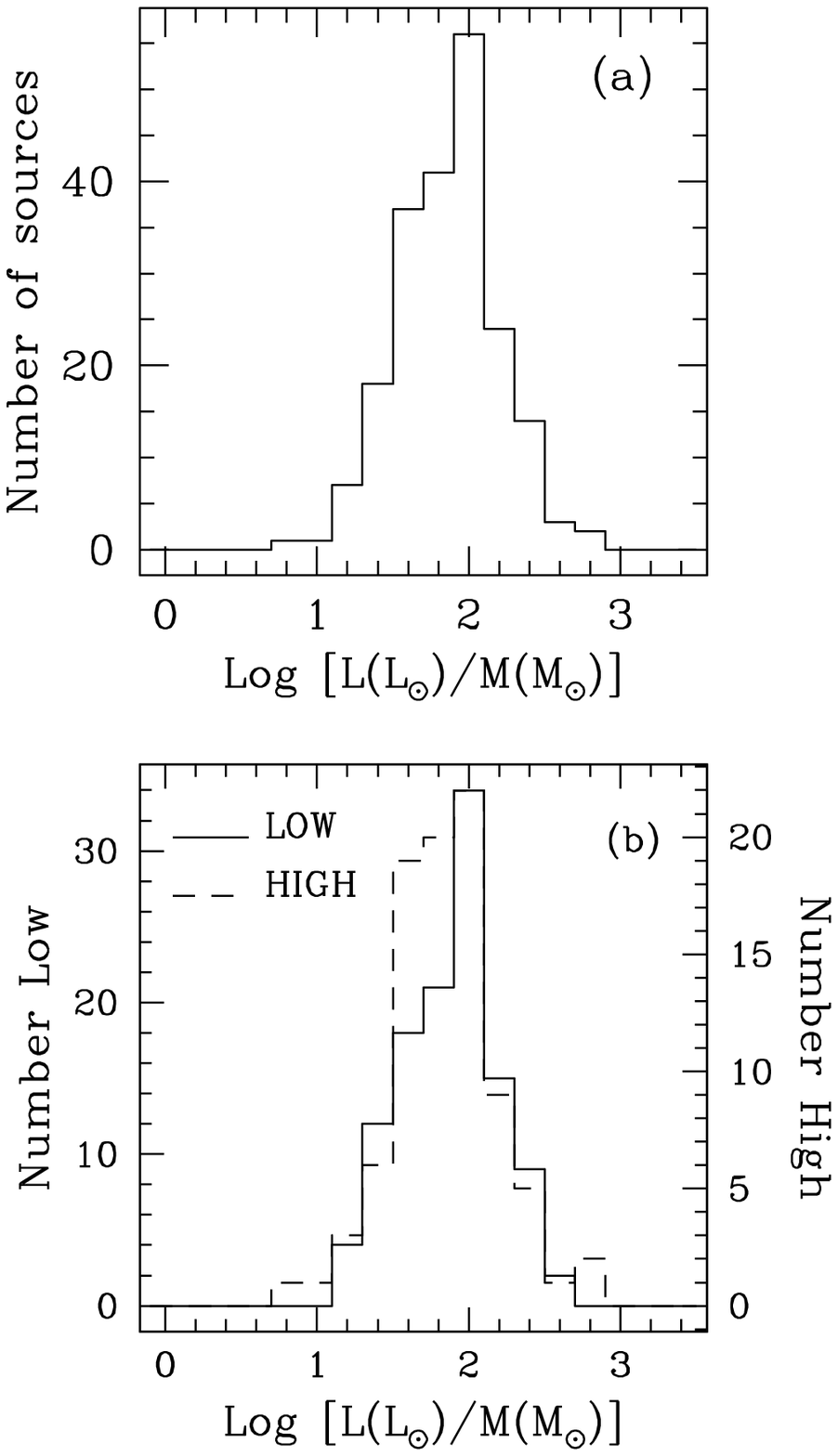}}
\caption{{\it (a)} Histograms of the distribution of the luminosity-to-mass ($L/M$) ratio for all the sources, 
where $M$ is the sum of the masses of clumps located $<90''$ from the nominal IRAS point source  position. {\it (b)} Same as above, for the {\it Low} (solid line) and the {\it High} sources (dashed line).} 
\label{histo_5}
\end{figure}

Figure~\ref{histo_4}d shows the histogram of the distribution of masses of the clumps. The mean value of the clump mass is
320~$M_\odot$, although the median mass is much lower, $102~M_\odot$. These values are in agreement with those of 330~$M_\odot$ and
143~$M_\odot$, respectively, found by Williams et al.~(\cite{williams04}) for the  68 high-mass protostellar candidates of the
Sridharan/Beuther sample (Sridharan et al.~\cite{srid02}; Beuther et al.~\cite{beuther02}) assuming the near kinematic distance.
Williams et al.~(\cite{williams04}) have derived their clump masses using the dust opacity coefficients of Ossenkopf \&
Henning~(\cite{ossenkopf94}), and $T_{\rm d}$ ranging from 30 to 60~K. Note that the average value of $5.0\times10^3~M_\odot$ given
by Fa\'undez et al.~(\cite{faundez04}) refers to the total mass in each region; i.e., the sum of the masses of all the clumps in
that region. If we take into account all the clumps in each region the average mass is 955~$M_\odot$, which is still five times
lower than the one derived by  Fa\'undez et al.~(\cite{faundez04}). This could be due to the fact that their sample contains
more massive stars, as already suggested by the bolometric luminosities of their targeted sources (see Section~\ref{lumi}). Hill et
al.~(\cite{hill05}) in their recent survey of massive star-forming regions harbouring methanol masers and/or radio continuum sources
have reported an average mass for their sample of $1.5\times10^3~M_\odot$, and a median value of $1.0\times10^3~M_\odot$, for a dust
temperature of 20~K. The average mass would be $\sim 0.9\times10^3~M_\odot$ and the median $\sim 0.6\times10^3~M_\odot$, for a dust
temperature of 30~K, which is the one that we have used for our estimates. Such values are still much higher than the ones of our
sample. This, again, could indicate that the YSOs in their sample are intrinsically more massive. Unfortunately these authors
do not report the bolometric luminosity of the sources, and we cannot corroborate this hypothesis.

Figures~\ref{histo_4}e and \ref{histo_4}f show the histograms of the H$_2$ volume density and surface
density of the clumps. These have been derived assuming that the clumps have spherical symmetry and a mean
molecular weight of $\mu=2.29m_{\rm H}$. The average values are $9.5\times10^5$~cm$^{-3}$ and
0.4~g\,cm$^{-2}$, respectively, and the median values are $4\times10^4$~cm$^{-3}$ and 0.14~g\,cm$^{-2}$,
respectively.

%figure 6
\begin{figure}
\centerline{\includegraphics[angle=0,width=8cm]{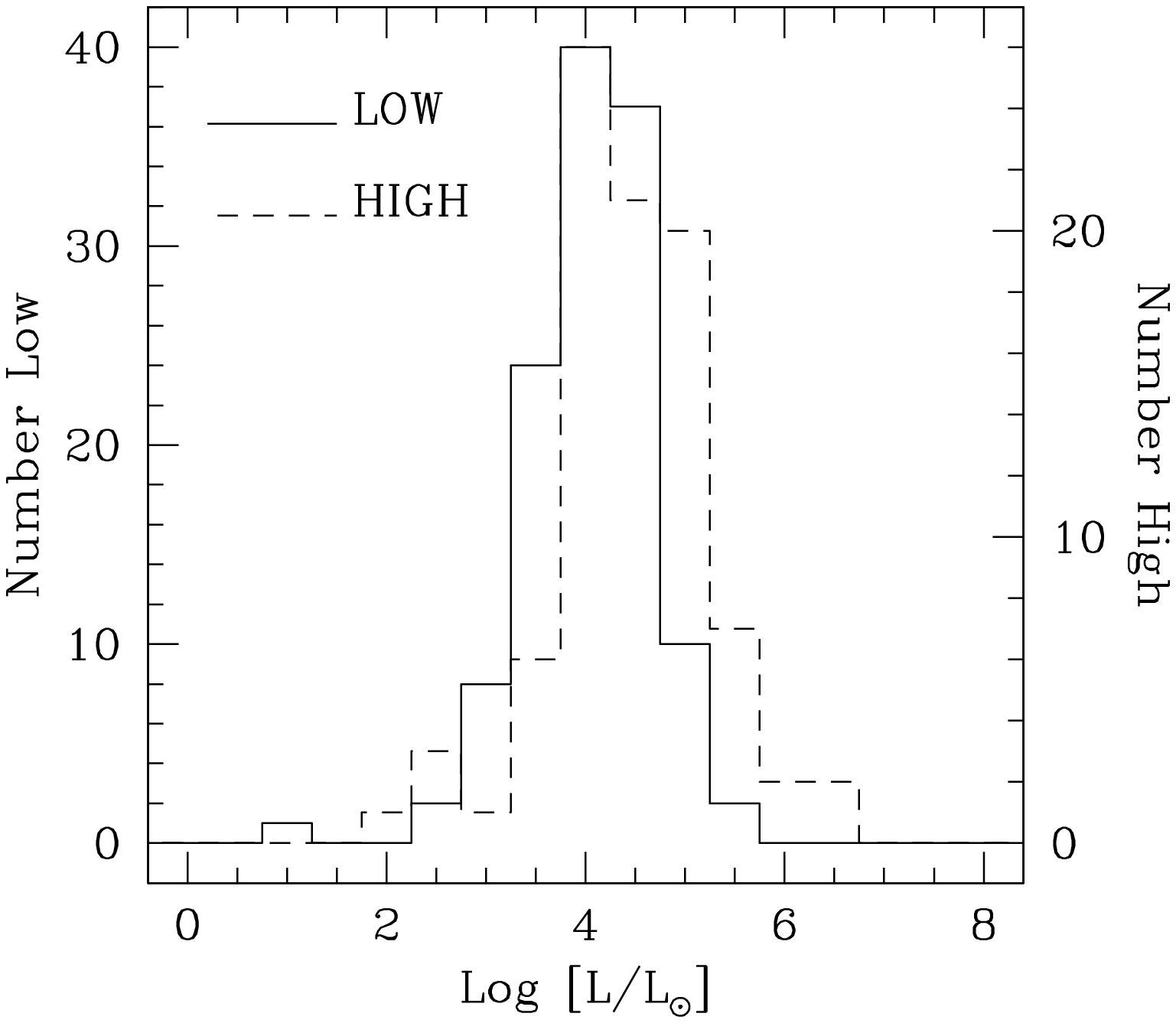}}
\caption{Histogram of the luminosity distribution calculated from the IRAS flux densities, for the {\it Low} (solid line) and the {\it High} sources (dashed line).} 
\label{histo_6}
\end{figure}

\section{Discussion}

\subsection{The Luminosity-Mass relation}
\label{lumass}

An important parameter for establishing the age of a core is the luminosity-to-mass ratio, $L/M$. This ratio is
expected to increase with time as more and more gas is converted into stars during the star formation process and
the embedded cluster becomes more luminous. Figures~\ref{histo_5}a  and \ref{histo_5}b show respectively the
$L/M$ ratio for all the sources in our sample, and for the {\it Low} and the {\it High} sub-samples separately.
The luminosity has been derived from the IRAS flux densities (see Sect.~\ref{lumi}), and  the mass is the sum of
the masses of the clumps located $<90''$ from the nominal IRAS point source  position (see Table~\ref{table_lumi}), as we
have considered only these clumps as associated with the IRAS source. The mean value of the distribution is
$99~L_\odot/M_\odot$ for all the sources, and  $98~L_\odot/M_\odot$ and $101~L_\odot/M_\odot$ for the {\it Low}
and the {\it High} sub-samples, respectively. These values are in agreement with the value of
$100~L_\odot/M_\odot$ found by Sridharan et al.~(\cite{srid02}) for the sources of the Sridharan/Beuther sample,
when rescaled to the Ossenkopf \& Henning~(\cite{ossenkopf94}) opacity of 1~cm$^2$\,g$^{-1}$ used by us to
derive  the masses. Sridharan et al.~(\cite{srid02}) report an average value of $0.05~M_\odot/L_\odot$, based on
the masses derived by Beuther et al.~\cite{beuther02}), which were estimated using the opacity of
Hildebrand~(\cite{hildebrand83}). This opacity is $\sim 2.5$ times smaller than the Ossenkopf \& Henning opacity
at 250~GHz. However, note that the masses estimated by Beuther et al.~\cite{beuther02}), as recently reported by
these authors, are wrong by a factor of 2 (see Beuther et al.~\cite{beuther05}); that is, they should be a factor
of 2 lower. Therefore, the average value of the $M/L$ ratio for their sample is actually $0.025~M_\odot/L_\odot$,
or $40~L_\odot/M_\odot$, which rescaled to the opacity of Ossenkopf \& Henning is $100~L_\odot/M_\odot$. These
values are slightly higher than the average value of $71~L_\odot/M_\odot$ reported by Fa\'undez et
al.~(\cite{faundez04}). However, as already mentioned, these authors have used the sum of the masses of all the
clumps in each region in their calculations. If we sum the mass of all the clumps in the regions, the average
value that we obtain is $77~L_\odot/M_\odot$ , a value similar to the one reported by those authors.

No significant difference is seen between the mean value of $L/M$ for the {\it Low} and the {\it High} sources. The same
conclusion has been found by Fontani et al.~(\cite{fontani05}) when comparing the {\it Low} sources of our sample with the {\it
High} sources of the sample of Sridharan/Beuther (Sridharan et al.~\cite{srid02}; Beuther et al.~\cite{beuther02}). These
latter authors have compared the $L/M$ values of the sources in their sample with those of known \UC\ regions with the same masses
(Hunter~\cite{hunter97}, Hunter at al.~\cite{hunter00}) and have found a $L/M$ ratio 5 times higher for the latter. Note that
all these masses have been wrongly calculated by these authors with a factor 2 higher (see Beuther et
al.~\cite{beuther05}). However, this error do not change the conclusions derived from their comparison, as both are affected by
the same factor. Mueller et al.~(\cite{mueller02}) and Fa\'undez et al.~(\cite{faundez04}) have found a similar result,
although the difference between cores associated with \UC\ regions and those not associated reported by these authors is not
statistically significant. Sridharan et al.~(\cite{srid02}) have suggested that the difference in $L/M$ ratio occurs because the
sources in their sample are in a younger pre-\UC\ phase, and that $L/M$ increases as the cores evolve and develop \UC\ regions.
Taking into account that the sources in our sample have $L/M$ ratios similar to those of the Sridharan/Beuther sample (see
Sect.~\ref{mass}), we conclude that they too are younger than \UC\  regions.

%figure 7
\begin{figure*}
\centerline{\includegraphics[angle=0,width=14cm]{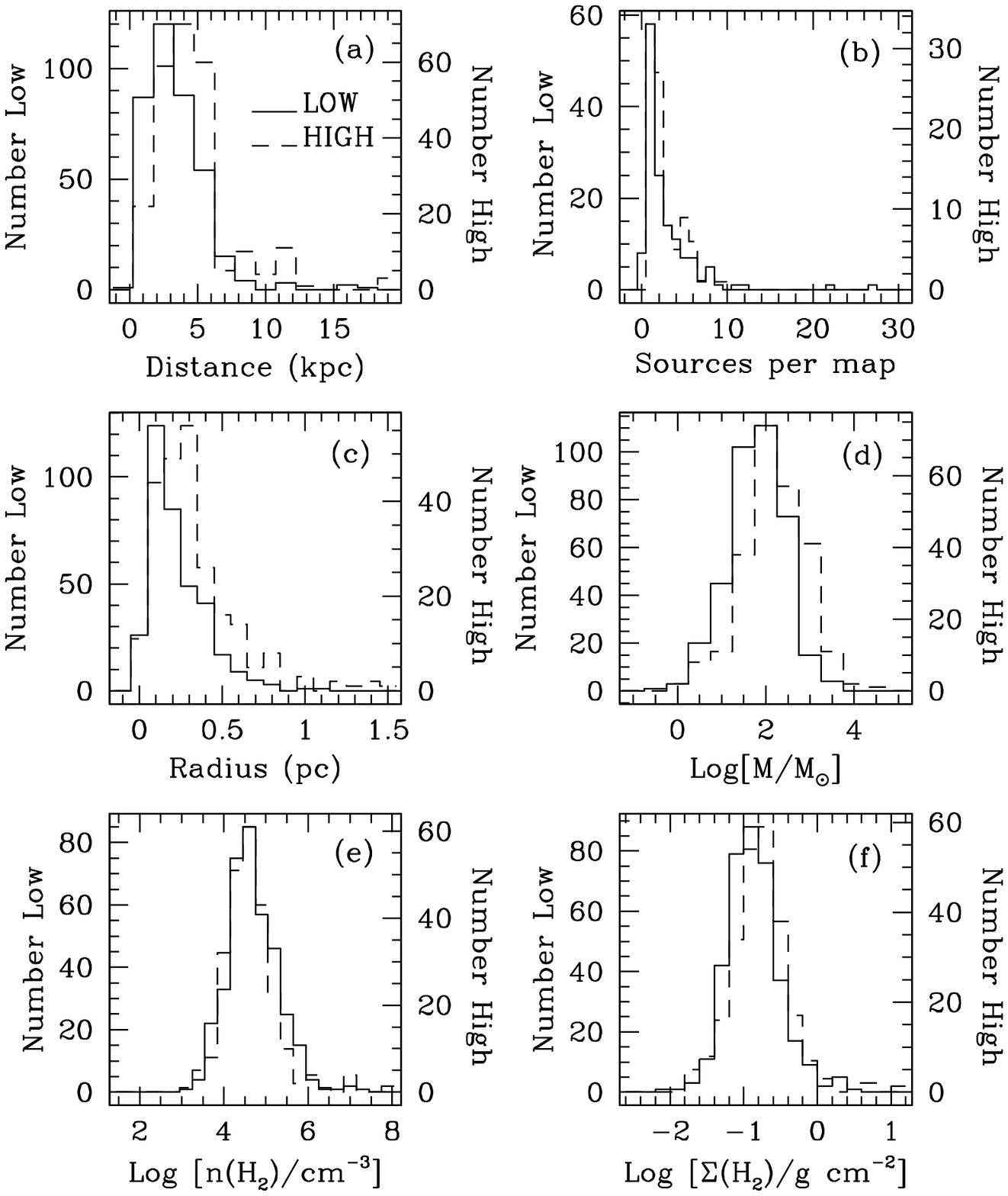}}
\caption{Histograms of {\it (a)} kinematic distance,  {\it (b)} number of clumps per map, {\it (c)} radius of the clumps,
{\it (d)} mass of the clumps, {\it (e)} H$_2$ volume density, and {\it (f)} H$_2$ surface density, for the {\it Low}
(solid line) and the {\it High} sources (dashed line). Note that in the regions for which  mosaiced observations were
carried out, we have included all the clumps in the field when computing the histograms, and not only the sources in a
field $900''\times 400''$, like for the other regions.} 
\label{histo_7}
\end{figure*}

\subsection{Low versus High}

One of the goals of this work is to carry out a comparative study of two sub-samples of massive YSOs, {\it Low}
and {\it High} believed to be in different evolutionary stages. To do this one should check whether the two groups
have different physical properties that could confirm a different evolutionary phase. Figure~\ref{histo_6} shows
the histogram of the luminosity distribution for both sub-samples, and Fig.~\ref{histo_7} shows comparative
histograms of some other physical properties, such as the number of clumps in each region, the size of the
clumps, their mass, H$_2$ volume density, and surface density. As can be seen in these figures, both sub-samples
{\it Low} and {\it High} are quite similar and do not show significant differences. This is the same conclusion
reached by Fontani et al.~(\cite{fontani05}) when comparing the properties (linewidth, luminosity, temperature,
NVSS-to-IRAS flux ratio) of the {\it Low} sources in our sample with the {\it High} sources of the
Sridharan/Beuther sample with luminosity $L<10^5 L_\odot$.

Analyzing the properties in more detail, one can see that the luminosity distribution of the {\it High}
sub-sample (Fig.~\ref{histo_6}) shows more sources at higher luminosities; the mean value for the {\it High}
sub-sample, $1.3\times 10^5 L_\odot$, is a factor of 5 higher than that for the {\it Low} sub-sample,  
$2.5\times 10^4 L_\odot$. The median value, which is $2.6\times 10^4 L_\odot$ for the {\it High} and $1.3\times
10^4 L_\odot$ for the {\it Low}, is only a factor of 2 different. The mean number of clumps per region is
similar for both sub-samples, 2.9 for the {\it Low} and 2.7 for the {\it High}, while the median number is 2.0 for
both sub-samples. The maximum number of clumps per region are 27 and 22 and correspond to the mosaics around the
{\it Low} sources 17225$-$3426 and 18014$-$2428, respectively. If one does not take into account these two
regions, the maximum number of clumps per region are comparable, with 12 clumps for the {\it Low} source
17141$-$3606, and 10 for  the {\it High} source 16085$-$5138. The morphologies of the clumps are quite similar in
both sub-samples (see Fig.~\ref{maps_mm}). Not surprisingly, the sources with fainter millimeter emission or no
emission at all towards the IRAS position, which are mostly {\it Low} sources, have no C$^{17}$O or CS detected (Fontani
et al.~\cite{fontani05}). 

Some differences are evident in the linear size and the mass  of the clumps, which have mean and median  values of
0.46~pc and 0.38~pc, and $163~M_\odot$ and $64~M_\odot$ for the {\it Low} clumps, and 0.67~pc and 0.56~pc, and
$561~M_\odot$ and $164~M_\odot$ for the {\it High} ones. Brand et al.~(\cite{brand01}) have found an opposite result for
the linear diameters of sources with $\delta \geq -30$, as they have found that clumps around {\it Low} sources are $\sim
3$ times larger than those around {\it High} clumps. However, those linear diameters have been obtained from molecular
line observations instead of from the dust continuum emission. 

As explained in Sect.~\ref{mass}, we have derived a
temperature of $\sim$30~K for all the clumps, with no significant difference between {\it Low} and {\it High} sources. We
have estimated the masses of the clumps adopting this value. However, our temperature estimates are based on a very
limited number of measurements, so that we cannot rule out the possibility that {\it High} sources are hotter than {\it
Low} ones. As a matter of fact, while  Fa\'undez et al.~(\cite{faundez04}) have found a temperature (32~K) similar to
ours for their sample of {\it High}-like sources, Sridharan et al.~(\cite{srid02}) determined a higher value (50~K) for a
similar sample. If one adopts 50~K also for our {\it High} sources, the mass estimates obtained by us should be
multiplied by 0.55 and the mean value of the mass of the clumps would be $309~M_\odot$, which is still a factor $\sim 2$
higher than the mean value of the {\it Low} sources. Therefore, the difference in mass of the clumps seems to be real.
Regarding the other physical properties, the H$_2$ volume density, and the surface density of the clumps, the average and
median values are $1.4\times10^6$~cm$^{-3}$ and $4.1\times10^4$~cm$^{-3}$, and 0.45~g\,cm$^{-2}$ and 0.13~g\,cm$^{-2}$
for the {\it Low} sources, and $2.6\times10^5$~cm$^{-3}$ and $3.9\times10^4$~cm$^{-3}$, and 0.29~g\,cm$^{-2}$ and
0.16~g\,cm$^{-2}$ for the {\it High} ones.

We have searched for possible asymmetries in the {\it Low} and {\it High} clumps by calculating ${\rm FWHM_x}/{\rm
FWHM_y}$, and we have found similar values for both sub-samples. The mean and median values are 1.04 and 0.95,
respectively, for the {\it Low} clumps, and 1.03 and 0.97 for the {\it High} ones. Such values indicate that the
clumps of both sub-samples are quite symmetric.

The two sub-samples seem to be in the same evolutionary stage, as suggested by the $L/M$ ratio (see
Sect.~\ref{lumass}). Therefore, in conclusion, there is no significant difference in the physical parameters
derived from the present millimeter continuum observations of the {\it Low} and {\it High} sub-samples with
$\delta < - 30\degr$. Both seem to be hosting high-mass (proto)stars with similar physical properties ($M$,
$n_{\rm H_2}$, radius, $L/M$) and evolutionary stage. However, in order to reasonably compare the results of our
southern study of {\it Low} and {\it High} sources with those obtained in previous northern hemisphere ($\delta
\geq -30\degr$) studies (Palla et al.~\cite{palla91}; Molinari et al.~\cite{moli96}, \cite{moli98a},
\cite{moli00}; Brand et al.~\cite{brand01}) which found differences between both sub-samples, more
observations at different wavelengths of line and continuum emission are still needed in the southern hemisphere.
In particular, it would be very important to properly estimate the temperature of the sources, either
through some  molecular  line observations, or though more infrared and (sub)millimeter observations to better
constrain their SEDs. Such temperature estimates could tell us whether the {\it Low} sources are
colder, and therefore younger, than the {\it High} sources, as observed in the northern hemisphere.

%figure 8
\begin{figure}
\centerline{\includegraphics[angle=0,width=8cm]{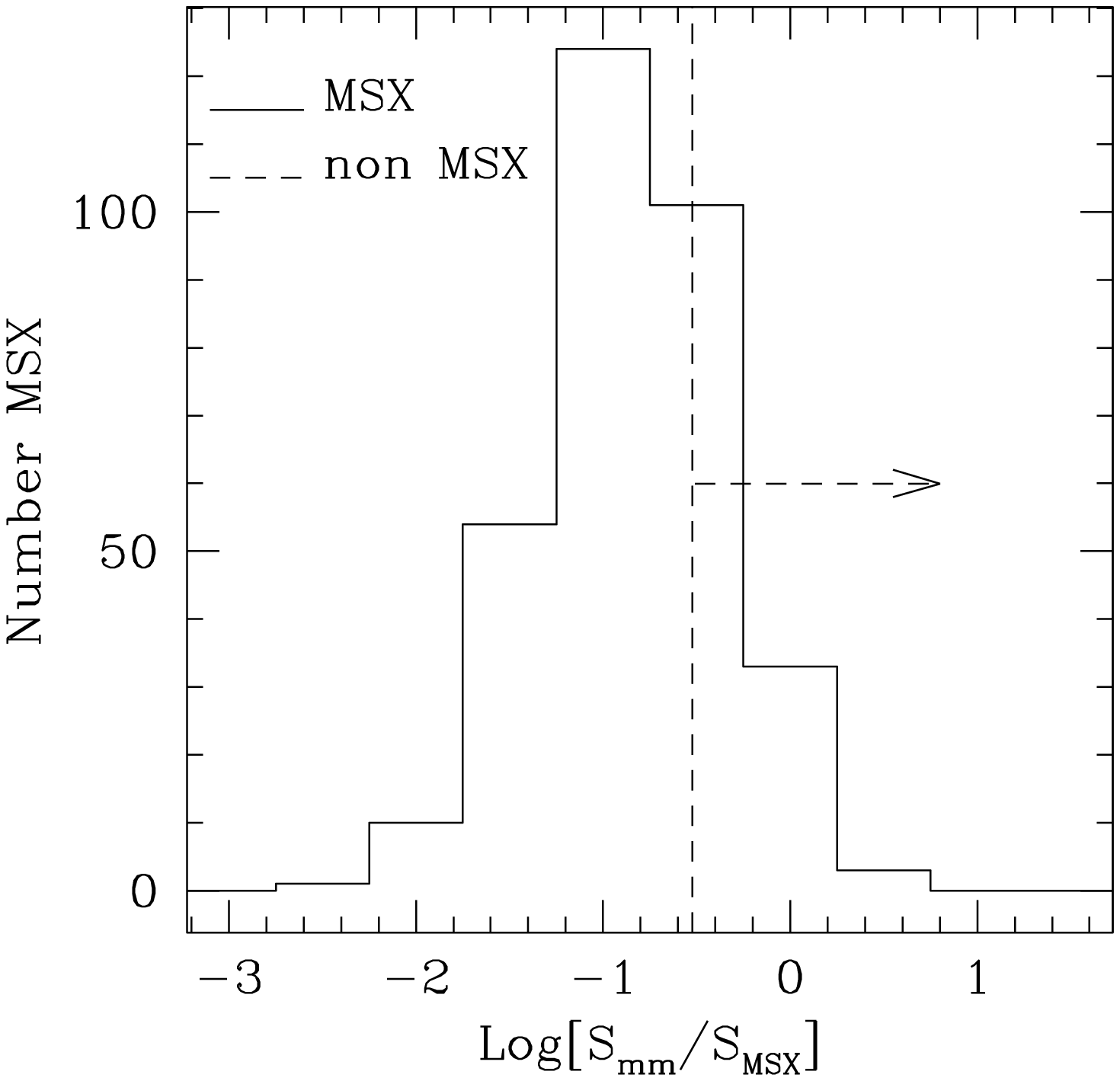}}
\caption{Histogram of the 1.2~mm to 21~$\mu$m integrated flux density ratio for clumps associated with point-like MSX emission (solid line).
The dashed line indicates the median value of $S_{\rm mm}/(1.5$~Jy), where $S_{\rm mm}$ is the 1.2~mm integrated flux density and
1.5~Jy is the MSX detection limit at 21~$\mu$m, for those clumps that are not associated with MSX emission. Note that due to the
fact that these clumps have not been detected at  21~$\mu$m, the median value of the ratio $S_{\rm mm}/(1.5$~Jy) is a lower limit.}  
\label{histo_8} 
\end{figure}

\subsection{MSX versus non MSX}

%figure 9
\begin{figure}
\centerline{\includegraphics[angle=0,width=8.7cm]{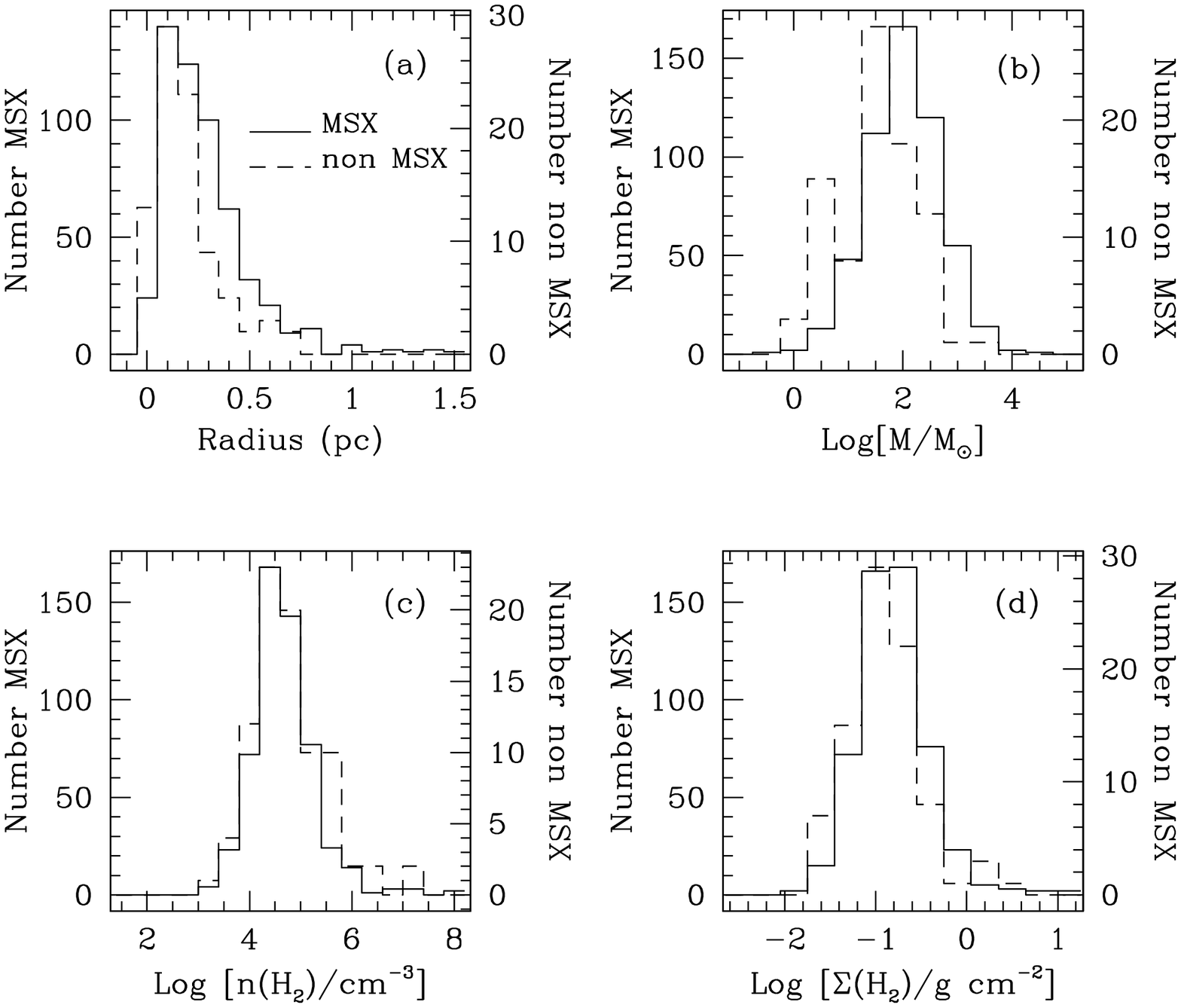}}
\caption{Histograms of {\it (a)} radius of the clumps, {\it (b)} mass of the clumps, {\it (c)} H$_2$ volume density,  and
{\it (d)} H$_2$ surface density, for clumps associated with MSX emission, either point-like or diffuse (solid line), and
those that are not (dashed line). Note that in the regions for which  mosaic observations were carried out, we
have included all the clumps in the field when computing the histograms, and not only the sources in a field $900''\times
400''$, like for the other regions.} 
\label{histo_9}
\end{figure}

As seen in the previous sections, the IRAS sources in our sample are usually associated with massive clumps likely
hosting high-mass (proto)stars. In an attempt to approach even closer to the earliest stages of high-mass star
formation, it would be of great interest to identify a core prior to the star formation process; namely a prestellar
or, better called, precluster core. Such a precluster core is expected to have density and size similar to those with
embedded high-mass protostars, but lower luminosity and temperature. The bulk of its luminosity is expected to be
emitted at millimeter and submillimeter wavelengths, with faint or no mid-IR or far-IR emission. One may reasonably
expect that such cores capable of forming massive stars but in a stage prior to the onset of star formation are
located close to massive luminous cores containing already formed high-mass (proto)stars. The millimeter continuum maps
have shown that there is usually more than one millimeter clump in the region surrounding the IRAS source, and that not
all of  them are associated with it; that is, they are located at $>90''$ from the nominal IRAS point source position. Therefore, we
have searched for precluster cores in the surroundings of the candidate massive (proto)stars by cross-correlating our
sample with the MSX Point Source Catalog, since the diameters of the clumps are comparable to the spatial resolution of
$18\farcs3$ of the mid-IR MSX observations. The cross-correlation has been done searching for mid-IR point sources  in
a radius of $40''$ around each millimeter clump, in any of the four bands (8~$\mu$m, 12~$\mu$m, 14~$\mu$m, or 21~$\mu$m)
in the MSX-PSC  version 6.0. In case that more than one millimeter clump was found at $\leq 40''$  from the same MSX point
source, we have arbitrarily chosen the nearest clump as the associated one. We have also compared by eye the millimeter
continuum maps with images of the MSX emission (see Fig.~\ref{maps_mm}), which has allowed us to identify millimeter
sources associated with mid-IR extended or diffuse emission. In the last column of Table~\ref{table_clumps} it is
indicated whether the clumps are associated with point-like or diffuse MSX emission,  or have no MSX emission at all.
The  positions in the MSX-PSC are accurate to an rms of $4''$--$5''$, and the calibration uncertainties are well within
13\% (Egan et al.~\cite{egan98}).

As a result of this study we have discovered 95  massive millimeter clumps not associated with mid-IR emission, meaning
no mid-IR point source at any of the four bands, nor diffuse emission, that are potential precluster clumps (see
Table~\ref{table_clumps}). However, this number could be a lower limit because the excess of mid-IR emission in the
observed regions produces, in some cases, confusion. Therefore, the association of some millimeter clumps with diffuse
MSX emission may not be real. In addition, there are 35 clumps in the MSX-PSC detected only at 8~$\mu$m, whose emission
could be due to contamination by Polycyclic Aromatic Hydrocarbons (PAHs) and not due to an embedded source. Until now
only a very limited number of such candidates have been found in the northern (Molinari et al.~\cite{moli98b}; Beuther et
al.~\cite{beuther02}; Sridharan et al.~\cite{srid02}), and southern hemisphere (Garay et al.~\cite{garay04}; Hill et
al.~\cite{hill05}). Note that in some cases, such as for example source 17221$-$3619, there are MSX sources located at
the edge of the millimeter cores, which might be either reflection nebulae or sources belonging to the same star-forming
region but in an older evolutionary stage: in fact, more evolved, and therefore hotter sources, are stronger emitters at
mid-IR wavelengths. Seventy out of the 95 clumps without MSX emission are clumps in maps around {\it Low} sources, which is
the 16\% of the {\it Low} clumps, while 20 (10\%) are clumps in maps around {\it High} sources.
Therefore, the {\it Low} sub-sample  contains in percentage more potential precluster clumps than the  {\it High}
sub-sample. 

%figure 10
\begin{figure*}
\begin{center}
\resizebox{13cm}{!}{\includegraphics{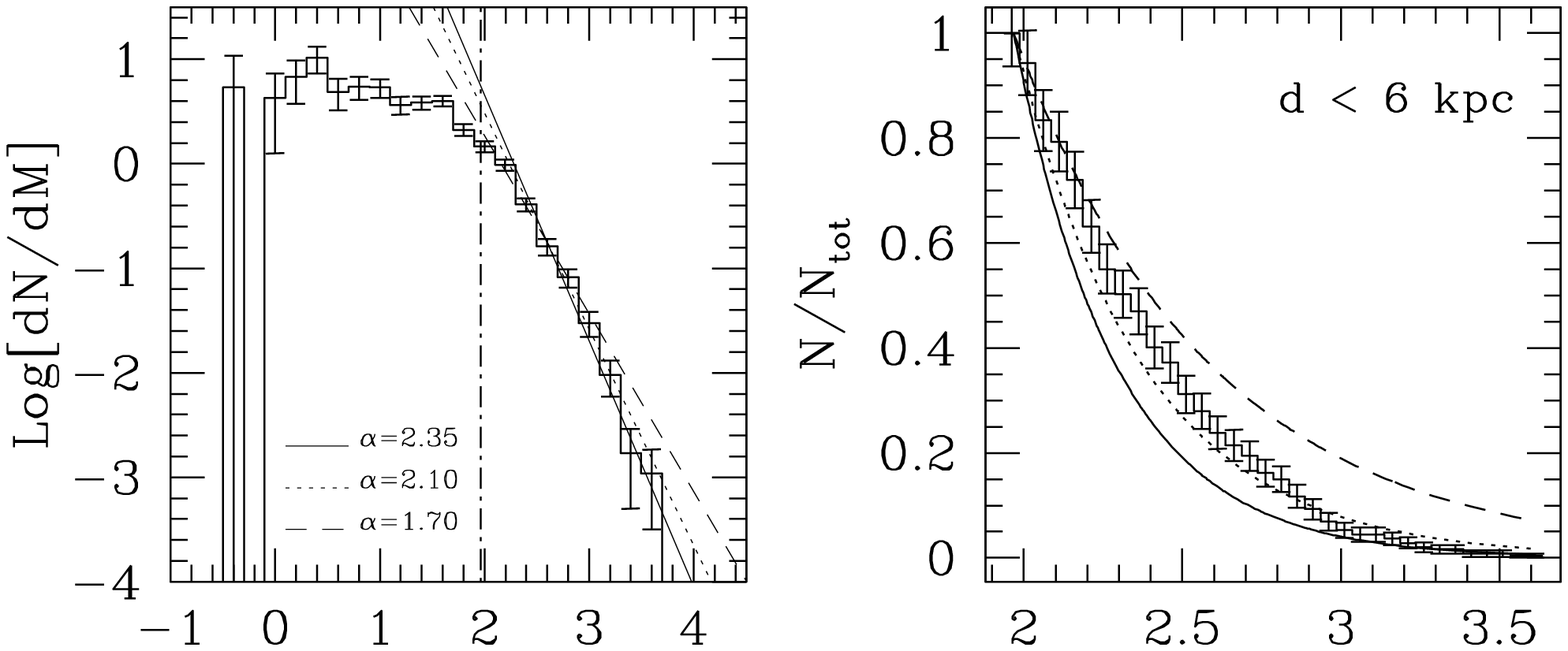}}
\\
\resizebox{13cm}{!}{\includegraphics{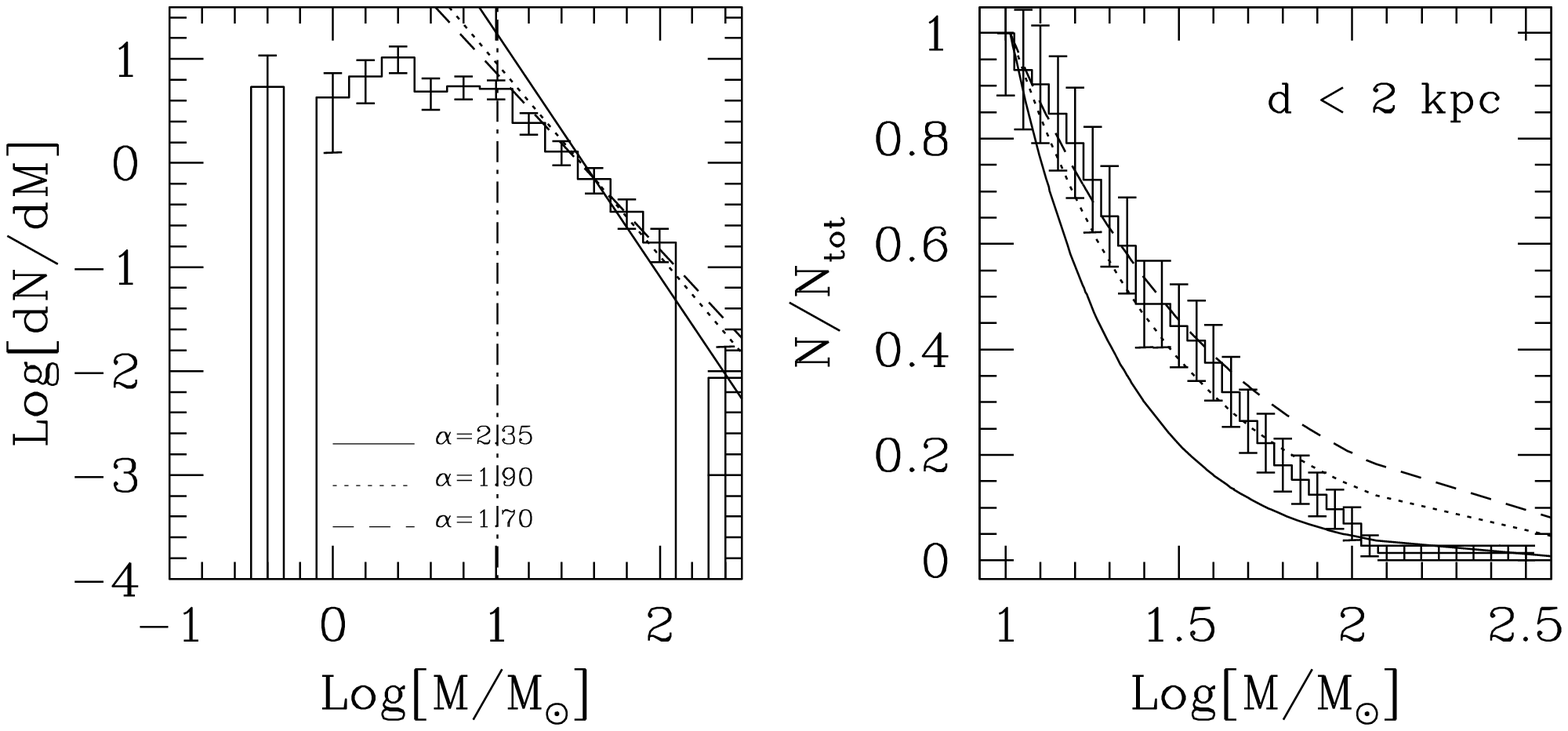}}
\end{center}
\caption{{\it Left top panel}: The mass spectrum of the 1.2~mm clumps detected at a distance $<6$~kpc. The solid line represents
the Salpeter IMF, $dN/dM \propto M^{-2.35}$; the dotted line is a $-2.1$ power law, obtained from the least square fit to the data, and the dashed line is a $-1.7$ power law. The
vertical dot-dashed line indicates the completeness limit at 6~kpc. {\it Right top panel}: The normalized cumulative mass
distribution of clumps with masses above the completeness limit at 6~kpc. The solid, and dashed lines are the same as in
the left panel, and the dotted line is a $-1.9$ power law, obtained from the least square fit to the data. {\it Left bottom panel}: Same as above for clumps detected at a distance $<2$~kpc. The vertical dot-dashed line
indicates the completeness limit at 2~kpc. {\it Right bottom panel}: Same as above for clumps with masses above the completeness
limit at 2~kpc.}
\label{imf}
\end{figure*}

Figure~\ref{histo_8} shows the histogram of the 1.2~mm to 21~$\mu$m integrated flux density ratio for the clumps associated  with
point-like MSX emission, where the flux density at 21~$\mu$m has been taken from the MSX-PSC.   The dashed line indicates the
median value of $S_{\rm mm}/(1.5$~Jy), where $S_{\rm mm}$ is the 1.2~mm integrated flux density and 1.5~Jy is the typical MSX
detection limit at 21~$\mu$m, for those clumps that are not associated with MSX emission. Note that because these clumps have
not been detected at 21~$\mu$m, the detection limit flux of 1.5~Jy is an upper limit for their emission at 21~$\mu$m, and
therefore, the median value of the ratio $S_{\rm mm}/(1.5$~Jy) is a lower limit. As can be seen in Fig.~\ref{histo_8}, there are 
relatively strong millimeter clumps that have no MSX counterpart, which indicates that they are not associated with mid-IR emission.
Therefore, some of the 95 millimeter clumps without MSX counterpart could be in fact precluster clumps, where luminous stars have
not yet formed.

Figure~\ref{histo_9} shows the comparative histograms of some physical properties, such as the radius, mass, H$_2$ volume
density, and  surface density of the clumps associated with MSX emission, either point-like or diffuse, and those not. As
can be seen in the distributions, the physical properties of the  two sub-samples are similar. Due to the fact that the
scale of the distributions is logarithmic, the properties would have to be very different in order to be clearly visible in
the histograms. The main difference is the mass of the clumps, which for non-MSX emitters is clearly lower, with a mean
value of $96~M_\odot$ and a median of  $33~M_\odot$ compared with the mean and median values of $336~M_\odot$ and
$106~M_\odot$, respectively, of the clumps with MSX counterparts. However, the dust temperatures of the clumps without
mid-IR emission could be significantly lower than that of the clumps with embedded massive (proto)stars, which is $T_{\rm
d}\sim 30$~K. In fact, Garay et al.~(\cite{garay04}) have found $T_{\rm d}< 17$~K for those clumps not associated with
MSX emission in their sample. Assuming a dust temperature of 15~K, the masses of the clumps should be corrected by a
factor $\sim 2.5$, and therefore, the masses of both MSX and non-MSX emitters would be more comparable. Regarding the other properties, the mean
values for the size, H$_2$ volume density, and  surface density are 0.3~pc, $4.7\times10^5$~cm$^{-3}$, and
0.2~g\,cm$^{-2}$ for the clumps without MSX counterpart (the densities maybe also scaled by a factor $\sim 2.5$ like the mass, in case
$T_{\rm d}= 15$~K), and 0.5~pc, $9.6\times10^5$~cm$^{-3}$, and 0.4~g\,cm$^{-2}$ for the clumps associated with mid-IR
emission. The median values are  0.3~pc, $4.0\times10^4$~cm$^{-3}$, and 0.11~g\,cm$^{-2}$ for the clumps without MSX
counterpart, and 0.4~pc, $3.6\times10^4$~cm$^{-3}$, and 0.14~g\,cm$^{-2}$ for those with MSX counterpart.

In order to assess whether the clumps without MSX emission are in an earlier evolutionary phase than the other
ones, it would be necessary to map the cores in different molecular tracers to better derive their physical
properties, and in particular their temperature, which should be lower for precluster cores.

\subsection{The Mass Spectrum of the clumps}

%\begin{table*}[bh]
\begin{table*}
\addtocounter{table}{2}
\caption[] {Power law indices of the mass spectrum of the clumps,  $dN/dM \propto M^{-\alpha}$, for different mass ranges. The
mass of the clumps has been derived either from the dust continuum or from  gas emission. The arrows indicate the whole mass
range for which the fit to the mass spectrum has been done.}
\label{table_imf}
\begin{tabular}{lccccc}
\hline
&\multicolumn{5}{c}{Mass Range $(M_\odot)$} \\
\cline{2-6} 
&\multicolumn{1}{c}{1.7--25} &
\multicolumn{1}{c}{10--35} &
\multicolumn{1}{c}{35--100} &
\multicolumn{1}{c}{100--$10^3$} &
 \multicolumn{1}{c}{$10^3$--$10^4$} 
\\
\hline
Dust emission \\
\hline
Tothill et al.\ (\cite{tothill02}) &---  &1.7 &--- &--- &--- \\
Beuther \& Schilke (\cite{beuther04}) &2.5 &--- &--- &--- &--- \\
Williams et al.\ (\cite{williams04})  & --- & \multicolumn{2}{c}{\leftarrowfill 1.14 \rightarrowfill} 
& \multicolumn{2}{c}{\leftarrowfill 2.32 \rightarrowfill} \\
This Paper  & --- & \multicolumn{2}{c}{\leftarrowfill 1.5--1.9 \rightarrowfill} 
& \multicolumn{2}{c}{\leftarrowfill 2.1 \rightarrowfill} \\ 
Reid \& Wilson (\cite{reid05}) & --- & \multicolumn{2}{c}{\leftarrowfill ~0.9 \rightarrowfill} 
& \multicolumn{2}{c}{\leftarrowfill  2.0 \rightarrowfill} \\ 
\hline	   
Gas emission \\
\hline
Kramer et al.\ (\cite{kramer98}) &--- &1.6--1.7 &1.6--1.8 &1.7--1.8 &1.7--1.8 \\
\hline 
\end{tabular}\end{table*}

Figure~\ref{imf} shows the histogram of the mass spectrum of the 1.2~mm clumps detected at a distance $d<6$~kpc ({\it top
panels}), and  $d<2$~kpc ({\it bottom panels}). The completeness limit, which has been estimated by calculating the mass of a
$5\sigma$ detection at the upper distance limit and is shown as a vertical dot-dashed line in the left panels, is $\sim
92~M_\odot$ at 6~kpc, and  $\sim 10~M_\odot$ at 2~kpc. The right panels show the normalized cumulative mass distribution of
clumps with masses above the completeness limit. The number of clumps above the completeness limit is 249 for $d<6$~kpc, and
79  for $d<2$~kpc. If the clump mass distribution can be represented by a power law of the type $dN/dM \propto M^{-\alpha}$,
then the histogram of the mass spectrum can be fitted with a straight line of slope $-\alpha$. The solid line in the figures
corresponds to $\alpha=2.35$, i.e., the Salpeter~(\cite{salpeter55}) IMF; the dashed line to $\alpha=1.7$, corresponding to
the mass function of molecular clouds derived from gas, mainly CO, observations (e.g.\ Kramer et al.~\cite{kramer98}); and the dotted
line corresponds to the slope computed from a least squares fit to our data. This value is $\alpha=2.10\pm0.02$, with a fit
correlation coefficient of 0.991, for the mass range $92~M_\odot\lesssim M \lesssim 4100~M_\odot$ ($d<6$~kpc), and
$\alpha=1.90\pm0.04$, with a fit correlation coefficient of 0.977,  for the mass range $10~M_\odot\lesssim M \lesssim
387~M_\odot$  ($d<2$~kpc). However, if ones does not take into account the last noisy point in the mass spectrum ($M =
387~M_\odot$) when computing the least squares fit, and fits the clumps detected at $d<2$~kpc with masses $10~M_\odot\lesssim
M \lesssim 120~M_\odot$, the slope of the fitted mass spectrum is much flatter, $\alpha=1.50$, with a fit correlation
coefficient of 0.999. Note that the errors reported for the slope values are those given by the fit, but do not represent the
true uncertainties in the values of $\alpha$, which depend also on other factors such as the binning used to
determine the mass spectrum.

As can be seen from Fig.~\ref{imf}, which shows the mass spectrum and the best-fit power law for two different mass
ranges, the mass spectrum, for masses above the completeness limit, cannot be fitted by a single power law for the whole mass
range; that is, for $10~M_\odot\lesssim M \lesssim 4100~M_\odot$. The clump mass spectrum is slightly steeper for the
high-mass end ($M \gtrsim 100~M_\odot$), and it flattens for lower masses ($M \lesssim 100~M_\odot$): for clumps detected at
$d<6$~kpc with masses above the completeness limit of $92~M_\odot$ ({\it top left panel}), the spectrum is well fitted with
$\alpha = 2.1$, consistent with a Salpeter-like power law; on the other hand, for clumps detected at $d<2$~kpc, with masses
$10~M_\odot\lesssim M \lesssim 120~M_\odot$, the mass spectrum is better fitted with $\alpha = 1.5$ or, if one takes also
into account a clump with a mass $M \lesssim 387~M_\odot$, with $\alpha = 1.9$, more consistent with the mass function of
molecular gas clouds  (e.g.\ Kramer et al.~\cite{kramer98}). 

Such a  change in the slope of the mass spectrum has been previously detected in other dust emission single-dish surveys of 
massive very YSOs (see Table~\ref{table_imf}). Williams et al.~(\cite{williams04}) have fitted the mass spectrum of their
high-mass sample with a very flat slope, $\alpha=1.14$ for $M < 100~M_\odot$, and with $\alpha=2.32$ for $M > 100~M_\odot$.
Reid \& Wilson~(\cite{reid05}) have fitted the submillimeter clump mass function in NGC~7538 with $\alpha=0.9\pm0.1$ for the
mass range $15~M_\odot\lesssim M \lesssim 100~M_\odot$, and with $\alpha=2.0\pm0.3$ for the mass range $100~M_\odot\lesssim M
\lesssim 2700~M_\odot$. In both cases, the breakpoint in the slope of the mass spectrum is well above the completeness limit
of their samples, which suggests that the change in slope is real and not an observational effect. A power-law with
$\alpha=1.7$ has been fitted by Tothill et al.~(\cite{tothill02}) to the dust condensations with masses in the range
$10~M_\odot\lesssim M \lesssim 35~M_\odot$  in the Lagoon Nebula, which is the same region that we have mapped in our mosaic
around 18014$-$2428 (Fig.~\ref{mosaic}). Assuming that the break in the mass spectrum for massive dust clumps is real and not
an artifact of the observations due to incompleteness, its shape is certainly interesting.

For $M \gtrsim 100~M_\odot$, the slope of the mass spectrum of the dust clumps is $\alpha=2.0$--2.32 (see
Table~\ref{table_imf}). Such massive clumps are associated with pre- and protoclusters, so that the mass
spectrum plotted in Fig.~\ref{imf} could correspond to the pre- and protocluster mass distribution.
Interestingly, a slope of 2--2.32 is similar to the clump mass distribution in low-mass star-forming regions,
which for masses $0.5~M_\odot\lesssim M \lesssim 10~M_\odot$ mimics the stellar IMF (e.g.\ Testi \&
Sargent~\cite{testi98}; Motte et al.~\cite{motte98}; Johnstone et al.~\cite{johnstone00},
\cite{johnstone01}). The similarity between our result and those obtained in low-mass star-forming regions
seems to suggest that the fragmentation of massive clumps may determine the IMF and the masses of the final
stars. In other words, the processes that determine the clump mass spectrum might be self-similar across a
broad range of clump and parent cloud masses. Reid \& Wilson~(\cite{reid05}), based on theories of molecular
cloud evolution (e.g.\ Gammie et al.~\cite{gammie03}; Tilley \& Pudritz~\cite{tilley04}), suggest that
turbulent fragmentation might be the dominant process that determines the shape of the clump mass spectrum.

For lower masses ($10~M_\odot\lesssim M \lesssim 100~M_\odot$) the slope of the mass spectrum is shallower
and consistent with the one found for clumps of similar masses observed in molecular lines ($\alpha\sim 1.7$;
Kramer et al.~\cite{kramer98}). This is also shallower than the spectrum of lower mass ($M \lesssim 10~M_\odot$)
pre- and protostellar dust clumps. However, such a comparison may be not appropriate because of the
different tracer used in the line and continuum surveys, as line observations are more sensitive to
low-density material than continuum imaging. Moreover, for our clumps the slope
of the mass spectrum in the interval $10~M_\odot\lesssim M \lesssim 100~M_\odot$ is quite uncertain, because
the mass range is relatively small and close to the completeness limit. Finally, the limited angular resolution of the single-dish observations of massive star-forming regions might be
insufficient to resolve the clumps into the cores associated with individual (proto)stars. This makes it hazardous to
compare the results with those of interferometric observations or with those of low-mass star-forming regions, which have
a much higher angular resolution. In fact, Beuther \& Schilke~(\cite{beuther04}) resolved 12 clumps, with masses
$1.7~M_\odot\lesssim M \lesssim 25~M_\odot$, in their interferometric study of the massive star forming
region IRAS~19410+2336 and determined a mass spectrum with a slope of 2.5, consistent with the stellar IMF.
More studies of this type are needed to establish the slope of the mass function in high-mass
star forming regions.

%Finally, it is hazardous to compare
%the results of single-dish with those of interferometric observations:
%the limited angular resolution of the single-dish observations of massive star-forming regions,
%might be insufficient to resolve the clumps into the cores associated with individual
%(proto)stars.

%This agreement suggests that the processes which shape the clump mass function are self-similar, at least across five orders of
%magnitude in clump mass. In other words, the overall shape of the mass function is independent of the total mass of the molecular
%cloud, over the range of cloud masses studied.

\section{Conclusions}

We have extended to the southern hemisphere the project started by Palla et al.~(\cite{palla91}) in the northern sky aimed at
identifying high-mass protostellar candidates. We have carried out a 1.2~mm  dust continuum emission survey with the bolometer
array SIMBA at the SEST antenna of a sample of 235 sources selected from the IRAS-PSC: 93 {\it High} sources with CS (Bronfman et
al.~\cite{bronfman96}), 125 {\it Low} with $\delta < -30 \degr$, observed in CS and/or C$^{17}$O (with the exception of
15579$-$5347) by Fontani et al.~(\cite{fontani05}), plus 17 {\it Low} with $\delta > -30 \degr$ observed in NH$_3$ (with the
exception of 18198$-$1429) by Molinari et al.~(\cite{moli96}).

Massive dust clumps have been detected in all but 8 regions, with usually  more than one clump per region. The dust emission shows
a variety of complex morphologies, sometimes with multiple clumps forming filaments or clusters. Most of the sources with faint
dust continuum emission are {\it Low} sources with no C$^{17}$O or CS detected towards the IRAS position. The mean clump has a
linear size of 0.5~pc, a mass of $320~M_\odot$ for a $T_{\rm d} = 30$~K, an H$_2$ density of
$9.5\times10^5$~cm$^{-3}$, and a surface density of 0.4~g\,cm$^{-2}$. The median values are 0.4~pc, $102~M_\odot$, 
$4\times10^4$~cm$^{-3}$, and 0.14~g\,cm$^{-2}$, respectively.

The mean value of the luminosity-to-mass ratio, $L/M$, is 99~$L_\odot/M_\odot$ for all the sources, and  $98~L_\odot/M_\odot$
and $101~L_\odot/M_\odot$ for the {\it Low} and the {\it High} sub-samples, respectively. Such values are $\sim 5$ times lower
than the average value of known \UC\ regions with the same masses. This  difference suggests that the sources in our sample,
both {\it Low} and {\it High}, are in a younger pre-\UC\ phase, as the $L/M$ ratio is expected to increase as the cores evolve
and develop \UC\ regions.

The physical properties of both {\it Low} and {\it High} sub-samples do not show significant differences (see Fig.~\ref{histo_7}).
The luminosity of the {\it High} sources is on average higher than that of the {\it Low} sources (see Fig.~\ref{histo_6}), as are
the linear diameters and the masses of the clumps. The similar  $L/M$ ratio for both sub-samples suggests that they are in a similar
evolutionary stage.

There are 95 massive millimeter clumps in the surroundings of the candidate massive (proto)stars that are not associated with
mid-IR MSX emission, either point-like or diffuse (see Fig.~\ref{maps_mm}).  70 out of the 95 clumps without MSX emission are
clumps in maps around {\it Low} sources, which is the 16\% of the {\it Low} clumps, while 20 are clumps in maps around {\it High}
sources, 10\% of the {\it High} clumps.  Such clumps are potential prestellar or precluster cores, as one may expect that the bulk
of their luminosity is emitted at millimeter and submillimeter wavelengths. The physical properties of these clumps are similar to
those of the MSX emitters, apart from the mass that is significantly lower than for clumps with MSX counterpart. However, such a
difference could be due to the potential precluster clumps having a lower dust temperature.

The mass spectrum of the clumps with masses above $M \sim 100~M_\odot$ has been fitted with a power-law index $\alpha = 2.1$,
consistent with the Salpeter~(\cite{salpeter55}) stellar IMF, $dN/dM \sim M^{-2.35}$, or with the low-mass pre- and protostellar dust
clumps mass spectrum. On the other hand, the mass spectrum for clumps with masses $10~M_\odot\lesssim M  \lesssim 120~M_\odot$ is
better fitted with a slope $\alpha = 1.5$, more consistent with the mass function of molecular gas clouds  (e.g.\ Kramer et
al.~\cite{kramer98}). For $M \gtrsim 100~M_\odot$, the massive dust clumps are probably tracing pre- or protoclusters, which have
a mass spectrum similar to the stellar IMF. This suggests a self-similar process which determines the shape of the mass spectrum over a
broad range of masses, from stellar to cluster size scales. For $M \lesssim 100~M_\odot$, the shallower slope of the mass
spectrum could be due to the limited angular resolution of single-dish observations, which is not
enough to resolve the clumps into their real star-forming entities.

\begin{acknowledgements}

It is a pleasure to thank the ESO/SEST staff for their support during the observations. We thank Robert Zylka for helping us with the
SIMBA data reduction, and for his suggestions that improved the quality of the reduction scripts that we used.

\end{acknowledgements}

\clearpage

%\begin{table*}[bh]
\begin{table*}
\addtocounter{table}{-3}
\caption[] {Observed IRAS sources, alongside with source type, H(igh) or L(ow) according to the classification of Palla et
al.~\cite{palla91}, position (precessed to J2000 coordinates), kinematic distance, and luminosity.  The last column shows the sum
of the mass estimated from the 1.2~mm continuum for clumps located at $<90''$ from the IRAS position. The luminosities and the masses
have been estimated assuming the near distance, $d_{\rm near}$, except where noted.}
\label{table_lumi}
\begin{tabular}{lccccccc}
\hline
&&\multicolumn{1}{c}{$\alpha({\rm J2000})$} &
\multicolumn{1}{c}{$\delta({\rm J2000})$} &
\multicolumn{1}{c}{$d_{\rm near}^{~\rm a}$} &
\multicolumn{1}{c}{$d_{\rm far}^{~\rm a}$} &
\multicolumn{1}{c}{$L^{\rm b}$} &
\multicolumn{1}{c}{$M$} \\
\multicolumn{1}{c}{IRAS name} &
\multicolumn{1}{c}{Type} &
\multicolumn{1}{c}{h m s}&
\multicolumn{1}{c}{$\degr$ $\arcmin$ $\arcsec$} &
\multicolumn{1}{c}{(kpc)} &
\multicolumn{1}{c}{(kpc)} &
\multicolumn{1}{c}{($\times 10^3 L_\odot$)} &
\multicolumn{1}{c}{($M_\odot$)}
 \\
\hline
08140$-$3559 & H & 08 15 59.0 & $-$36 08 18 &  3.8 &  3.8 & 10.6  & 95  \\
08211$-$4158 & L & 08 22 52.3 & $-$42 07 57 &  1.7 &  1.7 & 3.0  & 74  \\
08247$-$4223 & L & 08 26 27.6 & $-$42 33 05 &  1.4 &  1.4 & 1.5  & 17  \\
08438$-$4340 & H & 08 45 36.0 & $-$43 51 01 &  1.3 &  1.3 & 4.3  & 108   \\
08470$-$4243 & H & 08 48 47.9 & $-$42 54 22 &  2.2 &  2.2 & 15.2  & 294   \\
08476$-$4306 & H & 08 49 26.7 & $-$43 17 13 &  0.6 &  0.6 & 0.3 & 8  \\
08477$-$4359 & L & 08 49 32.9 & $-$44 10 47 &  1.8 &  1.8 & 3.3  & 100   \\
08488$-$4457 & L & 08 50 38.2 & $-$45 08 18 &	---$^{\rm c}$ &	---$^{\rm c}$ &  ---$^{\rm c}$  &   ---$^{\rm c}$\\ 
08563$-$4225 & L & 08 58 12.5 & $-$42 37 34 &  1.7 &  1.7 & 3.2  & 112   \\
08589$-$4714 & H & 09 00 40.5 & $-$47 25 55 &  1.5 &  1.5 & 1.8  & 40  \\  
09014$-$4736 & L & 09 03 09.8 & $-$47 48 28 &  1.3 &  1.3 & 3.6  & 20  \\
09026$-$4842 & L & 09 04 22.2 & $-$48 54 21 &  1.9 &  1.9 & 2.3  & 45  \\
09131$-$4723 & L & 09 14 55.5 & $-$47 36 13 &  1.7 &  1.7 & 2.6  & 118   \\
09166$-$4813 & L & 09 18 26.6 & $-$48 26 26 &  2.3 &  2.3 & 2.1  & 48  \\
09209$-$5143 & L & 09 22 34.6 & $-$51 56 23 &  6.4 &  6.4 & 13.6  & 736   \\  
09566$-$5607 & H & 09 58 23.3 & $-$56 22 09 &  6.8 &  6.8 & 37.6  & 545   \\
09578$-$5649 & H & 09 59 31.0 & $-$57 03 45 &  1.7 &  1.7 & 8.3  & 91  \\
10019$-$5712 & H & 10 03 40.5 & $-$57 26 39 &  1.8 &  1.8 & 4.5  & 62  \\
10038$-$5705 & H & 10 05 31.9 & $-$57 19 54 &  6.0 &  6.0 & 17.1  & 135   \\
10088$-$5730 & L & 10 10 38.7 & $-$57 45 32 &	---$^{\rm c}$ &	---$^{\rm c}$ &  ---$^{\rm c}$  &  ---$^{\rm c}$  \\
10095$-$5843 & L & 10 11 15.8 & $-$58 58 15 &  1.1 &  2.9 & 0.7 & 13  \\
10102$-$5706 & L & 10 12 03.7 & $-$57 21 26 &  0.8 &  2.9 & 0.4 & ~n.d.$^{\rm d}$   \\
10123$-$5727 & L & 10 14 08.8 & $-$57 42 12 &  0.9 &  3.0 & 2.5  & 36  \\
10156$-$5804 & L & 10 17 26.8 & $-$58 19 46 &	---$^{\rm c}$ &	---$^{\rm c}$ &  ---$^{\rm c}$  &  ---$^{\rm c}$  \\
10184$-$5748 & H & 10 20 14.7 & $-$58 03 38 &  5.4 &  5.4 & 297   & 4100  \\
10276$-$5711 & H & 10 29 30.1 & $-$57 26 40 &  5.9 &  5.9 & 71.6  & 1115  \\
10277$-$5730 & L & 10 29 35.4 & $-$57 45 34 &  5.8 &  5.8 & 32.4  & 156   \\
10286$-$5838 & H & 10 30 31.5 & $-$58 53 30 &  5.9 &  5.9 & 49.1  & 313   \\
10295$-$5746 & H & 10 31 28.3 & $-$58 02 07 &  5.0 &  5.0 & 681   & 3810  \\
10308$-$6122 & L & 10 32 39.8 & $-$61 37 33 &  1.2 &  ---  & 1.5  & 6  \\
10317$-$5936 & L & 10 33 38.1 & $-$59 51 54 &  8.9 &  8.9 & 57.5  & 452   \\
10320$-$5928 & H & 10 33 56.4 & $-$59 43 53 &  9.1 &  9.1 & 273   & 3690  \\
10337$-$5710 & H & 10 35 40.7 & $-$57 26 15 &  0.4 &  4.2 & 0.1 & 2  \\
10439$-$5941 & L & 10 45 54.0 & $-$59 57 03 &  2.6 &  2.6 & 40.9  & 491   \\
10501$-$5556 & H & 10 52 11.0 & $-$56 12 26 &  2.5 &  2.5 & 7.1  & 158   \\
10521$-$6031 & L & 10 54 11.0 & $-$60 47 30 &  8.1 &  8.1 & 40.3  & 288   \\
10537$-$5930 & L & 10 55 49.0 & $-$59 46 47 &  7.2 &  7.2 & 35.8  & 434   \\
10545$-$6244 & L & 10 56 32.9 & $-$63 00 34 &  2.0 &  ---  & 3.2  & ---$^{\rm e}$  \\
10548$-$5929 & L & 10 56 51.9 & $-$59 45 14 &  7.6 &  7.6 & 39.4  & 382   \\
10554$-$6237 & L & 10 57 25.0 & $-$62 53 10 &  3.0 &  3.0 & 3.2  & 63  \\
10555$-$6242 & H & 10 57 33.4 & $-$62 58 55 &  3.0 &  3.0 & 6.2  & 168   \\
10555$-$5949 & L & 10 57 37.5 & $-$60 05 32 &  8.6 &  8.6 & 3.2  & 149   \\ 
10559$-$5914 & H & 10 57 58.2 & $-$59 30 24 &  6.4 &  6.4 & 19.1  & 483   \\
10572$-$6018 & L & 10 59 19.3 & $-$60 34 10 &  7.2 &  7.2 & 46.7  & 366   \\
10575$-$5844 & L & 10 59 40.3 & $-$59 01 05 &	---$^{\rm c}$ &	---$^{\rm c}$ &  ---$^{\rm c}$  &  ---$^{\rm c}$  \\
\hline	   
\end{tabular}

   (a) Kinematic distance estimated from the NH$_3$ or CS line velocity using the rotation curve of Brand \& Blitz~(\cite{brand93}).\\    
   (b) Luminosities calculated by integrating the IRAS flux densities. \\   
   (c) No CS detected towards the IRAS source (Fontani et al.~\cite{fontani05}). \\
   (d) No source detected at millimeter wavelengths. \\
   (e) Millimeter sources located at $>90''$ from the nominal IRAS point source position.\\   
   (f) Far distance more than 200~pc from the galactic plane. \\
   (g) $d_{\rm near} < 100$ pc. Luminosity and mass estimated assuming $d_{\rm far}$. \\
   (h) No CS or C$^{17}$O observed. \\
   (i) Galactocentric distance $< 2$~kpc.
    	     
\end{table*}

\clearpage

\begin{table*}

\caption[] {Offset position, angular diameter, linear diameter, integrated flux density at 1.2~mm, mass, and density of the clumps
detected towards each IRAS source. The offset positions are relative to the nominal IRAS point source positions. The last column
indicates whether the clump is associated with MSX emission, either point-like or diffuse, or not. Linear diameters, masses, and densities of the
clumps have been estimated assuming the near kinematic distance when the distance ambiguity could not be resolved, except where
noted.}

\label{table_clumps}
\begin{tabular}{lcrrccccccc}
\hline
&&\multicolumn{1}{c}{$\Delta x$} &
\multicolumn{1}{c}{$\Delta y$} &
\multicolumn{1}{c}{$\theta$} &
\multicolumn{1}{c}{$D$} &
\multicolumn{1}{c}{$S_\nu{(\rm 1.2mm)}$} &
\multicolumn{1}{c}{$M_{\rm clump}^{\rm a}$} & 
\multicolumn{1}{c}{${n_{\rm H_2}}^{\rm a}$} \\
\multicolumn{1}{c}{IRAS name} &
\multicolumn{1}{c}{Clump} &
\multicolumn{1}{c}{($''$)}&
\multicolumn{1}{c}{($''$)} &
\multicolumn{1}{c}{($''$)} &
\multicolumn{1}{c}{(pc)} &
\multicolumn{1}{c}{(Jy)} &
\multicolumn{1}{c}{($M_\odot$)} &
\multicolumn{1}{c}{($\times 10^4$ cm$^{-3}$)} &
\multicolumn{1}{c}{MSX$^{\rm b}$}
 \\
\hline
08140$-$3559 &  1 &  $-32.0$  &  $ 16.0$    & 18.1    & 0.34  & 0.64 & 95  & 8.4 & P  \\
08211$-$4158 &  1 &  $ 0.00$  &  $-8.00$    & 48.7    & 0.41  & 2.45 & 74  & 3.7 & P  \\
08247$-$4223 &  1 &  $-40.0$  &  $ 16.0$    & 42.3    & 0.29  & 0.85 & 17  & 2.4 & P  \\
08438$-$4340 &  1 &  $-16.0$  &  $-64.0$    & 56.4    & 0.36  & 4.95 & 85  & 6.4 & D  \\
	     &  2 &  $-48.0$  &  $ 40.0$    & 35.8    & 0.23  & 1.32 & 23  & 6.7 & P  \\
08470$-$4243 &  1 &  $-8.00$  &  $-16.0$    & 34.6    & 0.36  & 6.15 & 294   & 20.6 & P  \\
08476$-$4306 &  1 &  $-8.00$  &  $ 8.00$    & 48.6    & 0.14  & 2.23 & 8  & 10.0 & P  \\
08477$-$4359 &  1 &  $ 24.0$  &  $-72.0$    & 35.6    & 0.31  & 1.80 & 58  & 6.8 & N  \\
	     &  2 &  $-24.0$  &  $-16.0$    & 34.4    & 0.30  & 0.93 & 30  & 3.9 & P  \\
	     &  3 &  $-24.0$  &  $-40.0$    & 12.5    & 0.11  & 0.39 & 13  & 34.2 & D  \\
08488$-$4457 &  1 &  $ 8.00$  &  $ 0.00$    & 29.9    & ---$^{\rm c}$& 0.4 & ---$^{\rm c}$ & ---$^{\rm c}$  & P  \\
08563$-$4225 &  1 &  $-8.00$  &  $-8.00$    & 45.4    & 0.36  & 3.50 & 97  & 6.9 & D  \\
	     &  2 &  $-176 $  &  $-136 $    & 27.9    & 0.22  & 0.81 & 22  & 6.8 & D  \\
	     &  3 &  $-216 $  &  $-136 $    & 28.4    & 0.23  & 0.62 & 17  & 4.9 & D  \\
	     &  4 &  $-32.0$  &  $-64.0$    & 19.6    & 0.16  & 0.56 & 15  & 13.5 & P  \\
	     &  5 &  $-80.0$  &  $-160 $    & 33.4    & 0.27  & 0.73 & 20  & 3.6 & P  \\
	     &  6 &  $-64.0$  &  $-104 $    & 25.1    & 0.20  & 0.37 & 10  & 4.3 & D  \\
08589$-$4714 &  1 &  $-8.00$  &  $-16.0$    & 34.1    & 0.24  & 1.79 & 40  & 9.3 & P  \\
	     &  2 &  $ 64.0$  &  $-112 $    & 11.1    & 0.08  & 0.35 & 8  & 53.3 & P  \\
09014$-$4736 &  1 &  $-64.0$  &  $ 0.00$    & 28.5    & 0.18  & 0.40 & 7  & 4.0 & P  \\
	     &  2 &  $-16.0$  &  $-16.0$    & 33.4    & 0.21  & 0.74 & 13  & 4.6 & P  \\
	     &  3 &  $ 136 $  &  $-128 $    & 18.8    & 0.12  & 0.29 & 5  & 9.9 & N  \\
	     &  4 &  $-136 $  &  $ 8.00$    & 18.3    & 0.12  & 0.25 & 4  & 9.3 & N  \\
09026$-$4842 &  1 &  $ 80.0$  &  $ 280 $    & 14.6    & 0.13  & 0.99 & 34  & 51.8 & P  \\
	     &  2 &  $ 344 $  &  $-136 $    & 9.85    & 0.09  & 0.39 & 13  & 65.9 & D  \\
	     &  3 &  $ 8.00$  &  $-16.0$    & 51.0    & 0.46  & 1.30 & 45  & 1.6 & P  \\
	     &  4 &  $ 384 $  &  $-144 $    & 13.7    & 0.12  & 0.27 & 10  & 17.3 & D  \\
09131$-$4723 &  1 &  $-24.0$  &  $ 24.0$    & 41.3    & 0.33  & 2.67 & 75  & 6.9 & P  \\
	     &  2 &  $-40.0$  &  $ 64.0$    & 26.1    & 0.21  & 0.90 & 25  & 9.2 & P  \\
	     &  3 &  $-56.0$  &  $-320 $    & 20.3    & 0.16  & 0.70 & 20  & 15.3 & N  \\
	     &  4 &  $-56.0$  &  $ 48.0$    & 20.2    & 0.16  & 0.66 & 18  & 14.4 & N  \\
	     &  5 &  $-152 $  &  $ 56.0$    & 28.5    & 0.23  & 0.38 & 11  & 3.0 & N  \\
09166$-$4813 &  1 &  $-80.0$  &  $-24.0$    & 32.2    & 0.35  & 0.93 & 48  & 3.7 & P  \\
09209$-$5143 &  1 &  $-312 $  &  $-32.0$    & 24.4    & 0.75  & 0.66 & 273   & 2.2 & P  \\
	     &  2 &  $ 32.0$  &  $ 16.0$    & 47.8    & 1.48  & 1.11 & 458   & 0.5 & D  \\
	     &  3 &  $ 16.0$  &  $-16.0$    & 31.4    & 0.97  & 0.67 & 278   & 1.0 & P  \\
09566$-$5607 &  1 &  $-8.00$  &  $-8.00$    & 34.3    & 1.13  & 1.16 & 545   & 1.3 & P  \\
09578$-$5649 &  1 &  $ 8.00$  &  $ 0.00$    & 46.2    & 0.37  & 1.63 & 45  & 3.0 & P  \\
	     &  2 &  $-16.0$  &  $-48.0$    & 50.1    & 0.40  & 1.65 & 46  & 2.4 & P  \\
10019$-$5712 &  1 &  $ 0.00$  &  $-8.00$    & 35.1    & 0.30  & 1.98 & 62  & 7.9 & P  \\
10038$-$5705 &  1 &  $-8.00$  &  $ 0.00$    & 8.23    & 0.24  & 0.37 & 135   & 33.1 & P  \\
10088$-$5730 &  1 &  $ 408 $  &  $-104 $    & 26.8    & ---$^{\rm c}$	& 0.72 & ---$^{\rm c}$   & ---$^{\rm c}$  & P  \\
	     &  2 &  $ 24.0$  &  $ 56.0$    & 39.8	& ---$^{\rm c}$   & 1.00 & ---$^{\rm c}$   & ---$^{\rm c}$  & P  \\
10095$-$5843 &  1 &  $-8.00$  &  $-8.00$    & 29.9    & 0.15  & 1.15 & 13  & 12.2 & P  \\
10102$-$5706 &~n.d.$^{\rm d}$ &    ---    &     ---     & ---     & ---   & ---  & ---   & ---  & P$^{\rm e}$  \\
\hline	   				        				    
\end{tabular}		        				    
    		     	                        					     
   (a) The mass of the clump, $M_{\rm clump}$, and the H$_2$ volume density, $n_{\rm H_2}$, are estimated from the 1.2~mm continuum for a dust temperature of 30~K, and using a gas-to-dust mass ratio of 100.\\		   
   (b) P: MSX point source within $40''$ from the millimeter clump peak emission position; D: diffuse MSX emission; N: No MSX emission. \\
   (c) No kinematic distance estimate.\\  
   (d) No source detected at millimeter wavelengths. \\
   (e) MSX point source within $40''$ of nominal IRAS position, or diffuse MSX emission. \\ 
   (f) Unresolved source. \\	     	                        					        
   (g) Linear size, mass and density estimated assuming the far kinematic distance. \\
   (h) The clumps listed have been detected  in a field offset by $4\farcm7$ from the nominal IRAS point source position. 
    		     	                        					     
\end{table*}

\clearpage

%figure 2
\addtocounter{figure}{-9}
\begin{figure*}
\centerline{\includegraphics[angle=0,width=10.7cm]{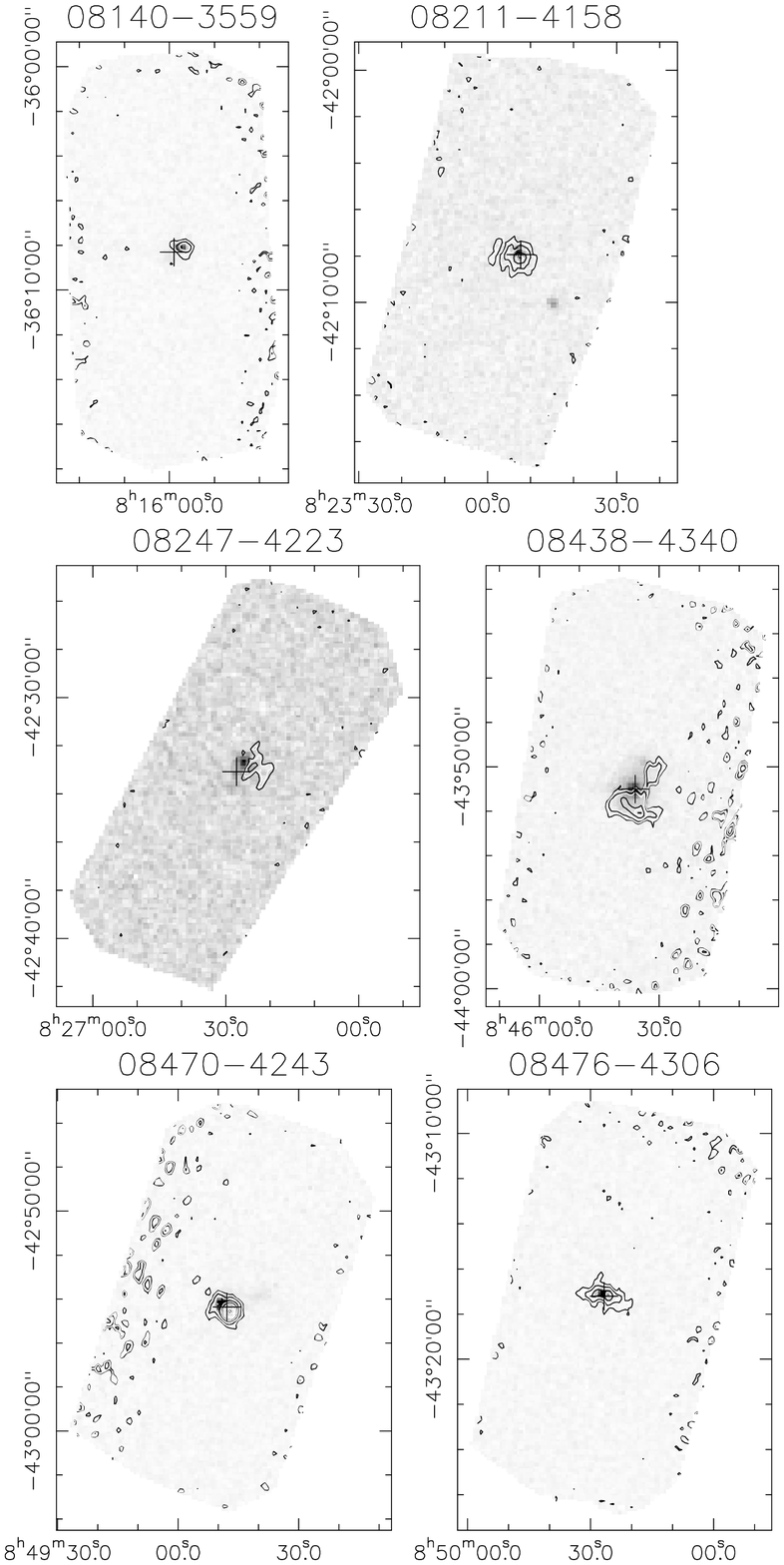}}

\caption{Overlay of the 1.2~mm continuum emission ({\it contours}) obtained with SIMBA at the SEST antenna, on
the MSX  emission at 21~$\mu$m (Band E) ({\it image}) towards the sources of our survey. The contour levels
are 3$\sigma$, 6$\sigma$, and from  6$\sigma$ to the maximum integrated emission in steps of 6$\sigma$, where $\sigma$ is the rms noise
of each map. The equatorial coordinates are in J2000 epoch. The noisy edges of the SIMBA map are visible. The
black cross marks the nominal IRAS point source position. For IRAS~16428$-$4109, the imaged field is offset by
$4\farcm7$ from the nominal IRAS point source position. For IRAS~18014$-$2428/18018$-$2426, the two IRAS
sources correspond to Mol~36 (IRAS~18014$-$2428; {\it black cross}) and Mol~37 (IRAS~18018$-$2426; {\it black triangle}) 
from Molinari et al.~(\cite{moli96}).}
\label{maps_mm}
\end{figure*}

\end{document}